\documentclass[superscriptaddress,twocolumn,preprintnumbers,amsmath,amssymb,floatfix,aps,prb]{revtex4-2}
\usepackage[english]{babel}
\usepackage[colorlinks=true, allcolors=blue]{hyperref}
\usepackage{graphicx,tabularx}
\usepackage[version=4]{mhchem} 
\usepackage{amssymb}
\usepackage{dcolumn}
\usepackage{rotating}
\usepackage{color}
\usepackage[mathcal]{euscript}

\definecolor{gray0}{gray}{0.0}
\definecolor{gray64}{gray}{0.25}
\definecolor{gray128}{gray}{0.5}
\definecolor{gray192}{gray}{0.75}
\definecolor{gray255}{gray}{1.0}

\begin{document}

\title{Predictive Inorganic Synthesis based on Machine Learning using Small Data sets: a case study of Hydrodynamic Diameter-controlled Cu Nanoparticles}
\newcommand{\AffQuATOMs}{UHasselt, Institute for Materials Research (IUMAT), Quantum \& Artificial inTelligence design Of Materials (QuATOMs), Martelarenlaan 42, B-3500 Hasselt, Belgium}
\newcommand{\AffHyMaD}{UHasselt, Institute for Materials Research (IUMAT), Hybrid Materials Design (HyMaD), Martelarenlaan 42, B-3500 Hasselt, Belgium}
\newcommand{\AffIMEC}{imec, IUMAT, Wetenschapspark 1, B-3590 Diepenbeek, Belgium}
\newcommand{\AffEnV}{Energyville, IUMAT, Thor Park 8320, B-3600 Genk, Belgium}
\newcommand{\AffDESINe}{UHasselt, Institute for Materials Research (IUMAT),Design and Synthesis of Inorganic Materials (DESINe), Martelarenlaan 42, B-3500 Hasselt, Belgium}
\newcommand{\AffELEC}{Electrochemistry Excellence Centre (ELEC), Materials \& Chemistry Unit, Flemish Institute for Technological Research (VITO), Boeretang 200, Mol, 2400 Belgium}
\newcommand{\AffLSMO}{Laboratory of Molecular Simulation (LSMO), Institut des Sciences et Ingénierie Chimiques, École Polytechnique Fédérale de Lausanne (EPFL), Rue de l’Industrie 17, CH-1951 Sion, Switzerland}

\author{Brent Motmans}
\affiliation{\AffQuATOMs}
\affiliation{\AffHyMaD}
\affiliation{\AffIMEC}
\affiliation{\AffEnV}
\author{Digvijay Ghogare}
\affiliation{\AffIMEC}
\affiliation{\AffDESINe}
\affiliation{\AffELEC}
\author{Thijs G.I. van Wijk}
\affiliation{\AffQuATOMs}
\affiliation{\AffIMEC}
\author{Joren Van Herck}
\affiliation{\AffLSMO}
\author{Saba Heidarian}
\affiliation{\AffQuATOMs}
\affiliation{\AffIMEC}
\author{Pieter De Meyer}
\affiliation{\AffDESINe}
\affiliation{\AffIMEC}
\affiliation{\AffEnV}
\author{Berend Smit}
\affiliation{\AffLSMO}
\author{An Hardy}
\affiliation{\AffDESINe}
\affiliation{\AffIMEC}
\affiliation{\AffEnV}
\author{Danny E.P. Vanpoucke}
\affiliation{\AffQuATOMs}
\affiliation{\AffIMEC}
\email[Corresponding author: ]{Danny.Vanpoucke@UHasselt.be}
\date{\today}

\begin{abstract}
Copper nanoparticles (Cu NPs) have a broad applicability, yet their synthesis is sensitive to subtle changes in reaction parameters. This sensitivity, combined with the time- and resource-intensive nature of experimental optimization, poses a major challenge in achieving reproducible and size-controlled synthesis. While Machine Learning (ML) shows promise in materials research, its application is often limited by scarcity of large high-quality experimental data sets. This study explores ML to predict the DLS-derived hydrodynamic diameter of Cu NPs from microwave-assisted polyol synthesis using a small data set of $25$ in-house performed syntheses. Latin Hypercube Sampling is used to efficiently cover the parameter space while creating the experimental data set. Ensemble regression models successfully predict hydrodynamic diameters with good predictive performance given the limited dataset. Since quantitative regression requires a unique DLS-derived hydrodynamic diameter, the regression model is restricted to mono-modal DLS distributions, while a complementary classification model identifies synthesis conditions for which quantitative prediction is applicable. Using equivalent out-of-sample validation (OOB and LOOCV), the ML and DoE models showed comparable generalization (MAE $= 40.92$ and $40.46$ nm, respectively). The final ensemble model achieved an $R^2$ of $0.74$, an MAE of $23.81$ nm, compared to an $R^2$ of $0.60$, an MAE of $33.54$ nm for the DoE model, while retaining the complete synthesis parameter space, making it better suited for synthesis guidance. Additionally, classification models using both random forests and Large Language Models (LLMs) are evaluated to distinguish between large and small particles. These classification models exhibited only modest predictive performance, indicating that this small dataset is insufficient to fully exploit the capabilities of complex LLMs. Overall, this study demonstrates that carefully curated small data sets, paired with robust classical ML, can effectively support the synthesis of Cu NPs and highlights that for lab-scale studies, complex models like LLMs may offer limited benefits. The validation experiments further indicate the potential of ML guided synthesis to improve the efficiency of experimental optimization and reduce resource consumption.
\end{abstract}

\keywords{machine learning, microwave-assisted polyol synthesis, small data sets, copper nanoparticles}

\maketitle

\section{Introduction}

Metal nanoparticles (NPs) play a central role in a broad range of applications, including catalysis, electronics, and biomedicine, due to their unique size- and shape-dependent properties.\cite{daniel:ChemInform2004,zhai:AdvancedFunctMat2010,gawande:ChemRev2016,pankhurst:JournalPhysD2003} However, controlling these properties requires very precise control of various NP characteristics, including their morphology (\textit{i.e.}, size and shape), composition, and surface chemistry down to the nanometer scale. Achieving this level of control remains one of the biggest challenges in NP synthesis, as small variations in the synthesis can lead to final particles with entirely different properties.\cite{gawande:ChemRev2016,baig:MatAdv2021} Traditionally, synthesis optimization relies on iterative, empirical approaches that are slow, resource-intensive, and often lack mechanistic insight.\cite{baig:MatAdv2021} To overcome these limitations, more systematic, data-driven strategies are an interesting route to explore. Indeed, recent advances in artificial intelligence (AI) and machine learning (ML) offer promising opportunities to accelerate the development of reproducible and tailored NP syntheses.\cite{baig:MatAdv2021,baum:JournalChemInfMod2021,butler:Nature2018, schmidt:NPJCompMat2019,sanchez-lengeling:Science2018,huang:Materials2023,agrawal:APLMaterials2016}

By uncovering hidden correlations in complex parameter spaces, ML can guide experimental design and reduce the number of required experiments.\cite{butler:Nature2018,schmidt:NPJCompMat2019,sanchez-lengeling:Science2018,huang:Materials2023} In materials science, ML is now routinely used to uncover the structure–property relationships of materials, thereby accelerating the discovery of new materials with specific properties.\cite{butler:Nature2018,schmidt:NPJCompMat2019,sanchez-lengeling:Science2018,huang:Materials2023} However, for synthesis optimization, the key challenge lies in predicting how synthesis parameters determine the resulting crystal structure or particle morphology. These synthesis–structure relationships are far less explored, yet crucial for rational synthesis design. As a result, recent data-driven approaches are shifting towards models that learn synthesis–structure relationships to accelerate synthesis optimization.\cite{agrawal:APLMaterials2016, vanpoucke:PolymerInt2022, williamson:InorgChem2023} Unlike traditional one-variable-at-a-time (OVAT) approaches, which are slow and fail to capture the multidimensional correlations that govern synthesis outcomes, ML can account for such interactions and thereby accelerate synthesis optimization.\cite{williamson:InorgChem2023,lah:EngAnalysisBoundElements2024} However, while large data sets underpin many ML successes \cite{Huo2022MLsynthesis,merchant-2023,curtarolo-2013}, they are rarely feasible in laboratory-scale chemistry, where the creation of each data point requires a significant amount of time and resources. In the specific case of inorganic NP synthesis, literature-based data sets are often inconsistent and biased, as failed experiments remain unpublished\cite{chen:AdvEnergyMat2020, raccuglia:Nature2016,ma:NPJCompMat2024}, as shown by Huo \textit{et al.}\cite{Huo2022MLsynthesis}, who expected up to $16$\% of the values to be incorrect. As such, robust ML workflows for small, high-quality data sets often remain underexplored.

ML can identify which synthesis parameters most strongly influence inorganic NP outcomes, as demonstrated by Guda \textit{et al.}\cite{guda:JournalPhysChemC2023}, who trained ML models on $30$ data points to classify the outcomes of gold NP synthesis (success, shape, and formation kinetics) based on the precursor concentrations, reducing agents, and surfactants, achieving classification accuracies of around $80$\%. ML also enables closed-loop experimentation, where ML models suggest subsequent experiments based on previous results, accelerating optimization and supporting high-throughput synthesis.\cite{tao:AdvFunctMat2021,epps:AdvMat2020} Pellegrino \textit{et al.}\cite{pellegrino:ScientificRep2020} trained neural networks (NNs), with $20$ initial data points, to predict the size, dispersion, and aspect ratio of \ce{TiO2} NPs as a function of temperature, pH, and precursor and shape-controller concentrations. They report mean relative errors of $17$\% for hydrodynamic radius, $34$\% for dispersion, and $19$\% for aspect ratio, which decrease to $9.1$\%, $13.9$\%, and $5.6$\% , respectively, after refinement with six additional data points. In a similar small-data setting, Braham \textit{et al.}\cite{braham-2019} applied support vector machines to classify and predict the thickness of \ce{CsPbBr3} nanoplatelets using $53$ experimental data points. Although the model achieved a high $R^2$ value ($0.97$), the corresponding RMSE ($0.756$) indicates a non-negligible prediction error relative to the limited range of the target variable ($2-7$ layers). Similarly, Williams \textit{et al.}\cite{williams:AppliedNano2023} combined Design of Experiments (DoE) and random forest models to accurately predict the yield of Ag nanowires in a polyol-based flow reactor, based on 31 samples, using temperature and reagent concentrations. They report prediction accuracies of up to $97$\% using random forest models, and an $R^2$ value of $0.85$ using regression.

In the current work, the prediction of copper (Cu) NP DLS-derived hydrodynamic diameter as function of synthesis parameters is explored. Cu NPs hold significant promise for various catalytic processes, such as electro-catalysis and photo-catalysis, due to their abundance, cost-effectiveness, and versatile properties.\cite{guo:AngewandteChemIE2014, reske:JACS2014, gawande:ChemRev2016} For example, Reske \textit{et al.}\cite{reske:JACS2014} showed that decreasing the size of Cu NPs significantly improves their catalytic activity. Different synthesis methods give rise to different types of Cu NPs, which may differ in particle size, morphology, and extent of oxidation. While metallic Cu NPs are nominally $Cu^0$, they often undergo (partial) surface oxidation depending on the synthesis conditions, leading to $CuO_x$ phases of varying composition.\cite{gawande:ChemRev2016} The wide variety of possible synthesis outcomes, as well as the large parameter space, complicates the development of robust synthesis procedures to obtain specific properties.

Though ML is a tool with great potential for Cu NP synthesis, two challenges must be addressed. The first challenge is finding a good quality data set to train the ML model.\cite{chen:AdvEnergyMat2020, raccuglia:Nature2016, ma:NPJCompMat2024} To tackle this, a consistent and qualitative data set is built in-house. The second challenge is overcoming the data hungry nature of typical ML methods. Since the goal is to make synthesis optimization more efficient using ML, relying on hundreds of preliminary experiments to generate training data would be impractical in a lab-scale setting, both in terms of time and resource consumption. Instead, we focus on constructing a carefully selected, information-rich dataset that maximizes experimental variation while minimizing experimental effort.\cite{vanpoucke:JournalApplPhys2020, vanpoucke:PolymerInt2022} Finally, within the context of lab-scale experiments, we emphasize interpretable ML: the model should not only predict outcomes, but ideally also provide insight into the role of different synthesis parameters.

In this work, we develop an interpretable ML model for Cu NP synthesis based on a small, high-quality data set, aiming to accelerate synthesis optimization while providing fundamental insights into parameter effects. Beyond improving synthesis optimization, reducing the number of required experimental iterations also has practical implications for resource efficiency. In laboratory-scale synthesis, minimizing the number of experiments directly reduces chemical consumption, energy usage, and waste generation, highlighting the broader relevance of data-driven approaches. In addition, performance of the interpretable ML model is also compared with the more classical DoE approach.

Recent work by Zheng \textit{et al.}\cite{zheng-2023}, which applies large language models (LLMs) to synthesis protocols reported in the MOF literature, served as inspiration to examine the potential of LLMs for synthesis prediction. Motivated by this, we investigate whether LLMs can provide meaningful predictive performance in a small-data experimental setting and benchmark their performance against a classic ML approach.

\begin{figure}[!tb]
  \includegraphics[width=\columnwidth]{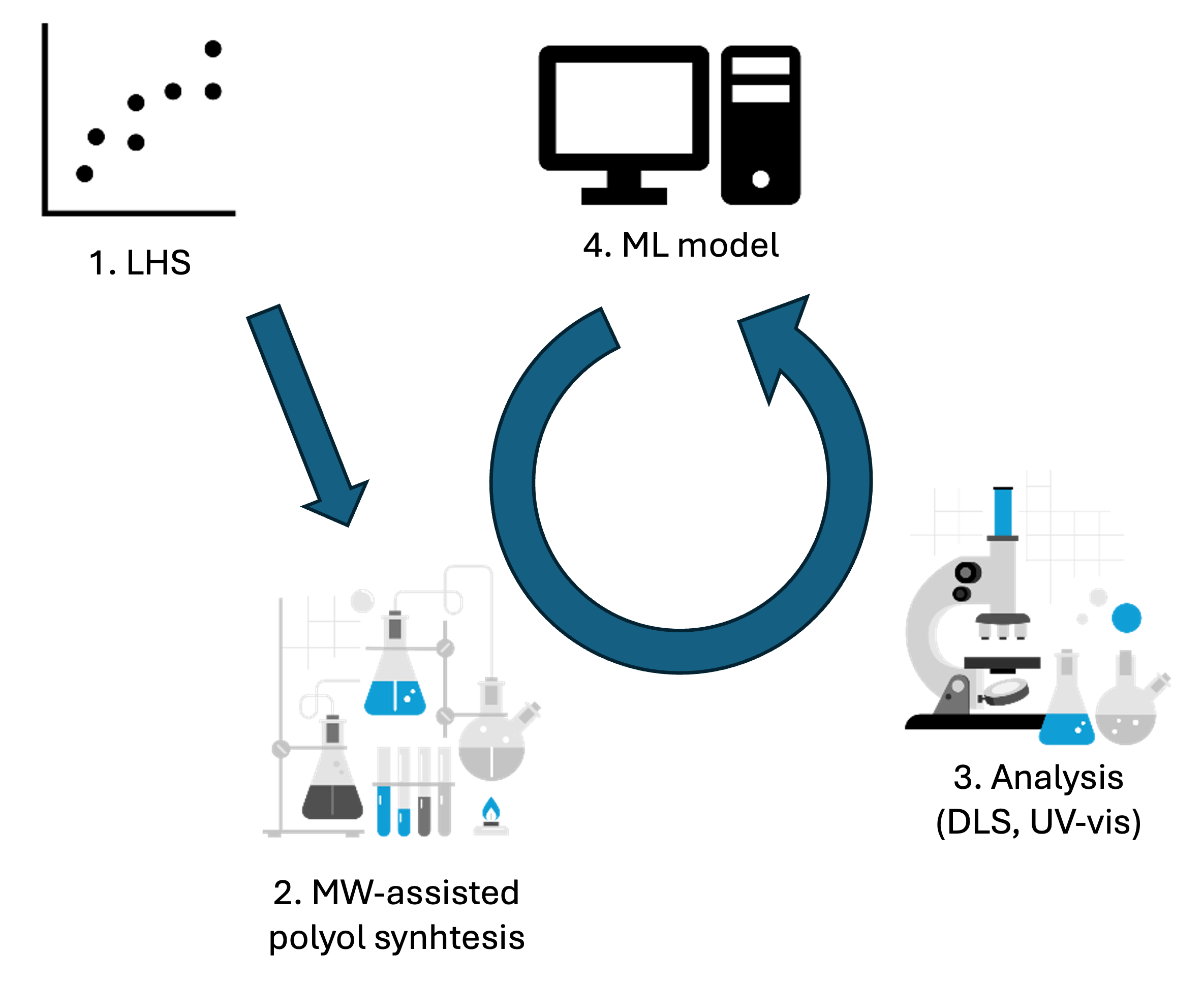}
  \caption{Visualization of the workflow followed to create the data set and model the syntheses.}
  \label{Fig:General workflow}
\end{figure}

\section{Methodology}

Within this work, we follow a four step workflow (Figure~\ref{Fig:General workflow}), of which the latter three can form a closed-loop experimentation methodology. The initial set of parameters for the data set is selected using Latin Hypercube Sampling (LHS) to achieve optimal parameter space coverage.\cite{borisut:Processes2023} The sampled syntheses are performed in the lab via MW-assisted polyol synthesis. The resulting dispersions are analyzed via dynamic light scattering (DLS) to determine the hydrodynamic diameter of the Cu NPs, while the surface plasmon resonance (SPR) peak is determined using UV-Vis measurements. The effective synthesis parameters\cite{fn-exp} and corresponding hydrodynamic diameters are combined into the data set for the ML modeling. The investigated parameter ranges were selected based on previous Cu polyol synthesis studies and practical considerations. The temperature range of $175$ to $200$ $^\circ$C was chosen based on the reported temperature-dependent reduction of Cu(II) in ethylene glycol under alkaline conditions\cite{teichert:DaltonTrans2018}, while the upper limit was chosen close to the boiling point of ethylene glycol and constrained by the operating specifications of the microwave reactor. The Cu precursor concentration was limited to $1–10$ mM to maintain a sufficiently broad but experimentally manageable parameter range.\cite{vreeland-2015} Reaction times of $1$ to $20$ minutes were selected based on the rapid kinetics reported for microwave-assisted Cu polyol synthesis.\cite{santanaotero-2025} A constant tetramethylammonium hydroxide (TMAH) to Cu molar ratio of 4:1 was used to maintain strongly alkaline conditions, motivated by the reported effect of a comparable hydroxide to Cu ratio on Cu(II) reduction.\cite{teichert:DaltonTrans2018} All predictive models developed in this work are trained using data collected within the same range of experimental synthesis conditions. Consequently, they are intended for interpolation within the investigated synthesis parameter space rather than extrapolation beyond the investigated ranges. In a closed-loop approach, the results of the ML models can be used to assist in further experimental syntheses, or further validation of the ML model.

\subsection{Experimental synthesis}

\subsubsection{Materials}

For all experimental syntheses, absolute ethanol ($\geq99.8$\% analytical grade) obtained from Fisher Chemical and copper(II) acetate monohydrate (\ce{Cu(OAc)2.H2O}, ACS reagent), tetramethylammonium hydroxide (TMAH, $25$ wt. \% in \ce{H2O}) and ethylene glycol ($\geq99$\%) obtained from Sigma Aldrich are used. All reagents are used as received without further purification.

\subsubsection{Sampling synthesis parameters}

LHS is used to obtain an optimized data set with minimal correlation between the various experiments while providing an optimal coverage of the parameter space under investigation.\cite{teichert:DaltonTrans2018,borisut:Processes2023,braham:DaltonTrans2020} In this work, the parameter space comprises three dimensions:

\begin{enumerate}
    \item Cu precursor concentration (1 mM to 10 mM)
    \item Reaction time (1 minute to 20 minutes)
    \item Reaction temperature (175 $^\circ$C to 200 $^\circ$C)
\end{enumerate}

Using LHS, 25 experiments are selected that fill this parameter space as optimally as possible. (\textit{cf.} SI Table I)

\subsubsection{Cu NPs synthesis}

The Cu NPs are synthesized via the microwave (MW)-assisted polyol route\cite{fievet:ChemSocRev2018, blosi:JournalNPResearch2010, teichert:DaltonTrans2018, carroll:JournalPhysChemC2011, torras:CrystalGrowth2021} in a CEM Discover SP MW reactor. First, the amount of \ce{Cu(OAc)2} is weighed into the reaction vial. It is then dissolved in 5 ml of ethylene glycol under sonication. After the Cu precursor has completely dissolved, the TMAH is added to the solution and further mixed and degassed under sonication. TMAH is consistently used at a molar ratio of $4:1$ relative to the Cu precursor. The reaction vessel is then placed in the microwave reactor, with the head space filled with inert argon up to a pressure of 5 bar, and is first stirred for $2$ minutes at $80 \ ^\circ$C to ensure complete dissolution of the precursor. Afterwards, the temperature is raised to the desired reaction temperature with continuous stirring and maintained for the desired reaction time. After the reaction time has passed, the reaction is quenched by actively cooling the vessel with compressed air. A brown-red suspension is obtained, which is typical for Cu NPs.\cite{blosi:JournalNPResearch2010} To collect the Cu NPs, the reaction mixture is diluted with $15$ ml of ethanol and centrifuged at $11000$ rpm for $10$ minutes. Subsequently, the supernatant is removed, and the resulting NPs are again dispersed in $20$ ml of ethanol.

\subsubsection{Characterization}

The SPR of the synthesized Cu NPs in the ethylene glycol dispersion is determined by UV-Vis measurements with the Agilent Cary 5000 spectrophotometer equipped with a dual-beam configuration. Measurements are performed in a quartz cuvette ($10$ mm path length) over a wavelength range of $300\text{-}800$ nm. Baseline correction is applied using the ethylene glycol solvent as reference.

The hydrodynamic diameter of the synthesized Cu NPs is determined via DLS measurements with a Zetasizer Nano ZS, equipped with a $633$ nm He–Ne laser. For these measurements, the ethanolic Cu dispersions are sonicated for at least 10 minutes. Subsequently, five drops of the NP dispersion are diluted with ethanol to a total volume of approximately $2\text{-}3$ mL, ensuring the disposable polystyrene cuvette is filled above the required optical path.

\subsection{Computational modeling}

Two modeling approaches are applied to the dataset: ensemble regression ML implemented in the AMADEUS framework,\cite{AMADEUS2020,vanpoucke:JournalApplPhys2020} and classical DoE.\cite{shahriari:ProceedingsIEEE2015} A quantitative comparison is performed using RMSE, MAE, and $R^2$ to evaluate the predictive performance of ML relative to DoE.

\subsubsection{Ensemble regression models}

\begin{figure}[!tb]
  \includegraphics[width=\columnwidth]{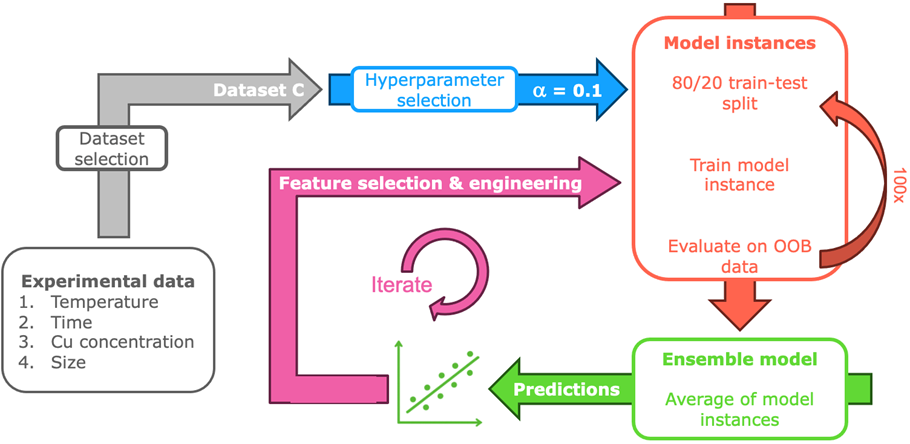}
  \caption{The computational modeling starts with selecting the best performing data set. Next, the optimal value for the regularization strength $\alpha$ is selected. Then, $100$ model instances are trained on random subsets of the data set. From these an ensemble model is created through averaging over all instances. Finally, through an iterative application of feature engineering and selection next generation models are created.}
  \label{Fig:Computational workflow}
\end{figure}

Since ML using small data sets is very sensitive to the selected training data, and hyperparameter tuning suffers from this volatility\cite{vanpoucke:JournalApplPhys2020, vanpoucke:PolymerInt2022}, we make use of the scikit-learn-based AMADEUS framework which was designed to deal with small data sets (Figure~\ref{Fig:Computational workflow}).\cite{vanpoucke:JournalApplPhys2020, vanpoucke:PolymerInt2022, pedregosafabianJournalMLResearch2011, AMADEUS2020} Pasting-type ensemble models consisting of $100$ base instances are created, using either linear or polynomial base models in combination with LASSO regularization.\cite{tibshirani:RoyalStatSocB1996}  Each of the instances is trained on a different randomly selected 80/20 split (in-bag/out-of-bag (OOB)) of the complete data set. The resulting final ensemble model is a single base instance equal to the average of the 100 base model instances.

To determine and compare the accuracy and generalization of the various models, root-mean-squared error (RMSE), mean absolute error (MAE) and coefficient of determination ($R^2$) are used as performance metrics. These errors can be determined on the in-bag and OOB data, with the difference between the two providing an indicator on model overfitting, while the individual values of the OOB error are quality indicators of the models' generalization. A low training error and high OOB error are very strong indicators of a model that is overfitting. In addition, performance metrics (RMSE, MAE, and $R^2$) are reported for predictions of the final ensemble model on the full dataset.

We first examine the impact of multi-modal data in our dataset. For this purpose, three data sets are created:

\begin{enumerate}
    \item Data set A: All $25$ data points are included in the data set, and the predicted hydrodynamic diameter is the smallest hydrodynamic diameter present in the DLS hydrodynamic diameter distribution. (\textit{cf.} SI Table II)
    \item Data set B: All $25$ data points are included in the data set, and the predicted hydrodynamic diameter corresponds to the peak with the highest percentage contribution in the DLS hydrodynamic diameter distribution. (\textit{cf.} SI Table III)
    \item Data set C: Only the $18$ data points with a mono-modal DLS hydrodynamic diameter distribution are included in the data set. (\textit{cf.} SI Table IV)
\end{enumerate}

For each of these data sets, ensembles of polynomial regression models of order $1$ to $8$ with LASSO regularization ($\alpha = 1.0$) are trained. The performance metrics of the best performing ensemble model are presented in Table~\ref{table:data set selection}. Since a quantitative regression model requires a uniquely defined target value, only data set C provides an unambiguous hydrodynamic diameter for every experiment. The improved predictive performance obtained for data set C therefore reflects the use of a well-defined regression target. Consequently, only data set C is considered further for quantitative regression model development. The trained regression models in this study thus focus on predicting the hydrodynamic diameter for mono-modal distributions (\textit{cf.} SI Figures 5, 6, and 7) and therefore is not intended to predict synthesis outcomes resulting in multi-modal DLS distributions. As a result, the regression model is applicable only within the mono-modal synthesis regime and should be combined with the accompanying modality classification model to identify synthesis conditions for which quantitative size prediction is appropriate.

\begin{table}[!tb]
  \caption{Performance metrics of the best-performing ensemble model for the different data sets.}
  \label{table:data set selection}
  \centering
  \begin{tabular}{lcccc}
    \hline
         & Polynomial degree & RMSE $[nm]$ & MAE $[nm]$ & $R^2$ \\
    \hline
    Data set A & $3$ & $63.46$ & $50.50$ & $0.42$ \\
    Data set B & $3$ & $59.44$ & $46.06$ & $0.35$ \\
    Data set C & $3$ & $34.09$ & $24.28$ & $0.73$ \\
    \hline
  \end{tabular}
\end{table}

When using a small data set, the growing number of features in polynomial models quickly leads to overfitting. Because the hyperparameter tuning of the regularization strength $\alpha$ (LASSO regularization) presented itself to be too volatile for automatic optimization, it is selected based on visual comparison of four values ($\alpha = 1.0, 0.1, 0.01,$ and $0.001$). Ensemble models with polynomials up to degree eight as base model are trained. Analysis of the ensemble model MAE shows an overall optimal performance for $\alpha = 0.1$ (\textit{cf.} SI Figure 1). Furthermore, the hyperparameter behavior shows no further improvement above polynomial degree three. Therefore, no features above degree three are considered, further reducing overfitting.

A further measure employed to reduce overfitting, is consideration of what one could call ensemble importance.\cite{vanpoucke:PolymerInt2022,VerdingP:AdvMaterTech2025} The AMADEUS framework tracks how often a polynomial feature is retained (\textit{i.e.} has non-zero coefficient) within the ensemble, after regularization of the features in each of the model instances. This provides a measure of importance of the features with regard to a well-generalizing model. A feature with an ensemble importance of $100$\% has a non-zero coefficient in all model instances in the ensemble. If a feature has an ensemble importance of $75$\%, this means that in $25$\% of the model instances in the ensemble, the coefficient for this feature is zero. Based on the trends in ensemble importance, seven features are selected (\textit{cf.} SI Figure 2):

\begin{enumerate}
    \item $[Cu]$
    \item $\textit{T}$
    \item $time$
    \item $[Cu]^2$
    \item $[Cu]^3$
    \item $[Cu] \times T$
    \item $[Cu] \times time$
\end{enumerate}

Since $[Cu]$, $\textit{T}$, and $\textit{time}$ are the base parameters used during synthesis, they are all retained as features. $[Cu]^2$ and $[Cu]^3$ are included as they play relatively important roles in both the second and third degree models. The interaction terms $[Cu] \times T$ and $[Cu] \times time$ are included, given their high importance in the second-degree polynomial; higher-order interaction terms are not included to limit model complexity and the probability of overfitting. Using these features, a linear regression model is trained, and iteratively, features are further engineered to obtain the optimal feature set.

\subsubsection{Classical statistical methods (DoE)}

Using DoE\cite{shahriari:ProceedingsIEEE2015} software JMP Pro (SAS Institute Inc.), a linear regression model is fitted for data set C, including terms up to the second degree. Insignificant terms are removed until only significant terms remain. In contrast to machine learning approaches, the available data set is not split into training and test data; all data points are used to fit the regression model. The performance of the final DoE model is determined by calculating the expected hydrodynamic diameter for all data points using the equation obtained.To obtain an out-of-sample performance estimate comparable to the OOB validation used for the ensemble ML models, the final DoE model was additionally evaluated using leave-one-out cross-validation (LOOCV). During each LOOCV iteration, one experiment was excluded from the dataset, the regression model was refitted using the remaining experiments, and the excluded experiment was subsequently predicted. The reported LOOCV performance metrics were calculated from the aggregated predictions of all left-out experiments.

\subsubsection{DLS-derived hydrodynamic diameter classification}

Unlike the regression model, the objective of the classification task is not to predict a quantitative DLS-derived hydrodynamic diameter but to classify the synthesis outcome into \textit{small} and \textit{large} DLS-derived hydrodynamic diameter categories. For this purpose, data set A is used in the classification task. (\textit{cf.} SI Table II) The aim is to predict whether the resulting particles will be \textit{large} or \textit{small} based on the synthesis parameters. Particles are defined as \textit{large} if they are larger than the median hydrodynamic diameter in data set A; if they are smaller than the median hydrodynamic diameter, they are defined as \textit{small}. For this classification task, the performance of complex LLMs is compared with that of a simple random forest classifier using accuracy and kappa ($\kappa$) as performance metrics. $\kappa$ indicates the agreement between predictions and actual labels, but corrects for coincidental agreements. It therefore indicates how well the model performs compared to random guessing. If $\kappa=1$ there is a perfect match between predictions and actual labels, while $\kappa=0$ indicates a performance equal to random guesses. If $\kappa<0$ the model performs even worse than random guessing.

Two different LLMs are trained and compared with each other for the binary classification task: GPT-J and LLAMA 3.1.\cite{jablonka:NatureMI2024, van-herck:ChemScience2024}, to predict the outcome as \textit{large} particles (`1') or \textit{small} particles (`0') based on the experimental input parameters. The LLMs are fine-tuned using the following prompt as input:

\begin{center}
    \texttt{‘What is the size of nanoparticles synthesized with a $<Cu \space conc>$ mol/litre Copper Acetate precursor concentration, a $<TMAH \space conc>$ mol/liter tetramethylammonium hydroxide concentration, heated for $<time>$ minutes at $<temperature>$ degrees celcius?’}
\end{center}

In this prompt, $<parameter>$ acts as a placeholder for the reaction parameter value. Because the TMAH concentration was maintained at a fixed molar ratio of $4:1$ relative to the Cu precursor concentration throughout this work, it does not constitute an independent synthesis variable. Nevertheless, it was retained in the prompt to provide a complete representation of the experimental synthesis protocol. The LLM predicts `0' or `1' corresponding to experimentally small or large NPs, respectively. The training data set is limited to a maximum of $15$ examples, while the test data set remained constant across all runs (\textit{e.g.}, $5$ examples). The number of epochs\cite{chollet-2025} is first evaluated at $10$, $25$, $50$, and $100$. For each combination of training size and epoch count, five independent runs are performed to compute average performance metrics. GPT-J is used for this screening. Because the task involves binary classification of a balanced data set, any accuracy above $50$\% indicates performance better than random guessing.

In comparison to the LLMs, an ensemble approach (\textit{e.g.}, random forest models) is used for binary classification. Automatic hyperparameter selection is used to determine the optimal value for the number of trees within a forest ($n\_estimators$), and the maximum depth of a tree ($max\_depth$). Based on this hyperparameter tuning, $n\_estimators$ is set to $200$ and $max\_depth$ to 3. With these hyperparameters, a single random forest is trained on a random subset of the data (80/20 train/test split). Repeating this process ten times, shows a high dependency on which specific data points are include in the training (and test) data, in line with previous observations with the volatility of results for small data ML.\cite{vanpoucke:JournalApplPhys2020} Our results show the models accuracy varying between $0.25$ and $0.75$. Because of this, an ensemble of $2500$ random forests is trained, each random forest on a different random subset of the data. Each generated model is tested on its OOB data, and these predictions are logged. After training and evaluating all $2500$ models, the OOB predictions of all models are used to obtain a final classification through hard voting. For the OOB predictions, the number of times they are classified as \textit{small} (`0') and the number of times as \textit{large} (`1') are counted for each data point. The class with the majority of votes will be the final predicted class.

\subsubsection{DLS distribution modality classification}

In addition to hydrodynamic diameter classification, a second classification task was introduced to predict whether the resulting hydrodynamic diameter distribution is mono- or multi-modal. The DLS analysis showed that both mono-modal and multi-modal distributions occur within the dataset. (\textit{cf.} SI Figures 5, 6, and 7) Furthermore, it was observed that the predictive performance of the regression model strongly depends on restricting the dataset to mono-modal samples (\textit{cf.} Table~\ref{table:data set selection}). Therefore, this classification task is included to explicitly account for this distinction.

The classification is based on the DLS peak structure, where samples exhibiting a single peak are labeled as (`0'), and samples with two or more peaks as (`1'). The resulting dataset is inherently imbalanced, with a majority of mono-modal samples. A support vector machine (SVM) classifier with a radial basis function (RBF) kernel is used for this task. Prior to training, all features (Cu precursor concentration, temperature, and reaction time) are standardized to ensure proper scaling.

Hyperparameter tuning is performed using a grid search over the regularization parameter (C) and kernel coefficient ($\gamma$), evaluated using leave-one-out cross-validation (LOOCV) due to the limited dataset size. The optimal hyperparameters are selected as $C = 1.0$ and $\gamma = 1.0$, based on the average classification accuracy (\textit{cf.} SI Figure 8). Although some hyperparameter combinations yield an apparent accuracy of $1.0$, this is not considered reliable in a LOOCV setting with a limited dataset (\textit{e.g.} SI Figure 9 always predicts the mono-modal class), where individual data points can strongly influence the result. Therefore, C $= 1.0$ and $\gamma = 1.0$ were selected as a representative set of hyperparameters based on the overall model behavior. Here, the classification accuracy is defined as the fraction of correctly classified samples, averaged over all LOOCV folds.

To assess model robustness in the small-data regime, an ensemble approach is adopted. A total of 100 SVM model instances are trained on different random stratified 80/20 train/test splits of the dataset. Due to the limited dataset size, model performance is sensitive to the specific choice of training data. Therefore, aggregating multiple model instances improves robustness. After training, each model is used to predict the full dataset, and the final classification for each data point is obtained through majority voting over all ensemble members.

\section{Results and discussion}

\subsection{Cu NP syntheses}

The DLS measurements show a peak above $5000$ nm for five samples. However, these peaks are not included in the analysis. Due to their very low intensity ($<1.5\%$), it can be assumed that these results are due to contamination, such as dust particles, during the measurements. The synthesis yields Cu NPs with a hydrodynamic diameter between $40$ nm and $350$ nm. The DLS shows a bi-modal and tri-modal hydrodynamic diameter distribution for five and two samples, respectively. The remaining samples show a mono-modal hydrodynamic diameter distribution. (\textit{cf.} SI Figures 5, 6, 7, and 16) Analysis of the DLS hydrodynamic diameter plots allows us to identify the smallest hydrodynamic diameter present. The median of these hydrodynamic diameters is $157.9$ nm over all samples. (\textit{cf.} SI Figure 3)

Based on the SPR location, the effective formation of Cu metal ($Cu^0$) NPs is verified for the different samples. The different UV-Vis spectra show a clear SPR peak between $590$ nm and $630$ nm for most samples, which is a strong indication of the presence of $Cu^0$ NPs.\cite{blosi:JournalNPResearch2010} Samples $4$, $6$, and $19$ do not show a clear SPR feature, which may be due to low particle concentration, oxidation\cite{vargas-urbano:Materials2022}, or the presence of ultra small particles ($<5 nm$), which do not exhibit a well-defined SPR; for the other samples, at least some $Cu^0$ is present.

Given the limited yield of the syntheses, no X-ray diffraction measurements could be performed to provide a conclusive answer, but the combination of the red-brown color of the solution and the presence of a clear SPR peak in the UV-Vis spectra (\textit{cf.} SI Figures 11 and 12, and SI Table V) strongly indicate that $Cu^0$ NPs have indeed formed. To validate this conclusion, Selected Area Electron Diffraction analysis is conducted on sample 22 using a transmission electron microscope operated at an accelerating voltage of $120$ kV. The resulting diffraction rings are indexed to the face-centered cubic structure of copper, with prominent rings corresponding to the (111), (200), and (220) planes. (\textit{cf.} SI Figure 4) The primary diffraction rings are assigned the following $d-$spacings: $2.08$ \AA \space for the (111) plane, $1.83$ \AA \space for the (200) plane, and $1.29$ \AA \space for the (220) plane. These values agree with the standard values for FCC copper (JCPDS card no. 04-0836)\cite{icdd2001}, confirming the formation of $Cu^0$ NPs.

As the focus of this work is not on a novel NP system, but on establishing an ML-guided workflow for synthesis optimization, a well-established Cu NP synthesis from the literature was selected.\cite{fievet:ChemSocRev2018,carroll:JournalPhysChemC2011,teichert:DaltonTrans2018} While more extensive structural characterization (\textit{e.g.}, TEM or XRD) would provide additional insight, these techniques are not strictly required for the present study, which focuses on hydrodynamic diameter as the primary descriptor for ML-guided synthesis. The chosen characterization methods are therefore sufficient to evaluate the objectives of this work.

\subsection{Quantitative DLS-derived hydrodynamic diameter prediction}

\begin{table*}[!tb]
  \caption{Performance metrics of different regression model generations.}
  \label{table:model generations performance}
  \centering
  \resizebox{\textwidth}{!}{
  \begin{tabular}{lcccccccc}
    \hline
         &  & \multicolumn{2}{c}{Training} & \multicolumn{2}{c}{OOB} & \multicolumn{3}{c}{Final ensemble model} \\
         & \# Features & RMSE $[nm]$ & MAE $[nm]$ & RMSE $[nm]$ & MAE $[nm]$ & RMSE $[nm]$ & MAE $[nm]$ & $R^2$ \\
    \hline
    1st generation & $19$ & $22.17$ & $16.05$ & $86.52$ & $68.51$ & $32.64$ & $22.53$ & $0.75$ \\
    2nd generation & $7$  & $29.40$ & $20.47$ & $51.85$ & $41.36$ & $32.61$ & $22.28$ & $0.75$ \\
    3rd generation & $11$ & $27.78$ & $19.79$ & $63.66$ & $52.18$ & $33.07$ & $23.35$ & $0.74$ \\
    4th generation & $6$  & $30.57$ & $22.39$ & $49.90$ & $40.92$ & $33.03$ & $23.81$ & $0.74$ \\
    DoE            & $2$  & $--$    & $--$    & $48.22^{a}$ & $40.46^{a}$   & $40.91$ & $33.54$ & $0.60$ \\
    \hline
  \end{tabular}}
  \raggedright
  \footnotesize
  $^{a}$ For the DoE model, these values were obtained using LOOCV.
\end{table*}

\subsubsection{Ensemble regression}

Ensemble regression models are created to predict Cu NP hydrodynamic diameter based on the synthesis parameters. As discussed above, this regression is restricted to mono-modal DLS distributions to ensure a uniquely defined quantitative target. Different generations of models are trained to investigate the influence of different features on model performance.

As first generation ensemble model, we consider the complete third-degree polynomial model with LASSO regularization ($\alpha = 0.1$). (\textit{cf.} SI Figure 13) This model has an $R^2$ of $0.75$, and RMSE and MAE of $32.64$ nm and $22.53$ nm, respectively. (\textit{cf.}, Table~\ref{table:model generations performance}) Despite the good performance of this first generation model, its main disadvantage is the use of an extensive set of $19$ features. With our data set consisting of only 18 samples, the risk of overfitting is significant. In addition, a larger feature set results in a more complex model and, therefore, higher computational cost for inference. With this in mind, the following generations are aimed at reducing the model size drastically to improve generalization and reduce computational cost.

In the second generation, the complexity of the model is strongly reduced by using only the seven features ($[Cu]$, $time$, $\textit{T}$, $[Cu]^2$, $[Cu]^3$, $[Cu] \times time$, $[Cu] \times T$) selected based on the ensemble importance analysis. The final ensemble model shows no significant loss in performance compared to the first generation (\textit{cf.}, Table~\ref{table:model generations performance}), despite the significant reduction in the number of features used. In addition, the training and OOB errors of the ensemble show that the second generation is more generalizable as their difference becomes smaller. The training error of the second generation increases slightly (MAE of $20.47$ nm compared to $16.05$ nm), while the OOB error decreases drastically (MAE goes from $68.51$ nm to $41.36$ nm). The smaller difference between training and OOB error indicates that the model is overfitting less and generalizes better. (\textit{cf.} SI Figure 14)

In the third generation model (\textit{cf.} SI Figure 15), the feature set is expanded with four additional features, aimed at the investigation of the influence of non-polynomial features (based on domain knowledge) on the model performance.

\begin{enumerate}
    \item $e^{[Cu]}$ is added because $[Cu]$ often has a high and consistent ensemble importance in the investigation of different polynomial models where it occurs in higher powers.
    \item $\frac{[Cu]}{time}$ can be interpreted as a measure of the amount of cations available per unit of time during the growth of the particles.
    \item $ln([Cu] \times time)$ provides a measure of the total exposure of the reduction medium to Cu ions.
    \item $\frac{[time]}{T}$ as a temperature-normalized reaction time. Although this parameter does not represent the classical thermal budget ($time \times \textit{T}$), it provides a simple heuristic descriptor that captures part of the interplay between time and temperature in a single scalar.
\end{enumerate}	

This more complex feature set, gives rise to a slight decline in the performance of the model due to stronger overfitting, leading to poorer generalization. (\textit{cf.}, Table~\ref{table:model generations performance}) Given the small difference in performance between this and the previous generation, a comprehensive analysis of the ensemble and dominance importance\cite{Azen:PsyMeth2003, Ceder:ChemMater2022} of the features is carried out. Based on this analysis features are selected, aimed at a good balance between sufficient complexity for good predictive quality and a reduced size while retaining good generalization, which resulted in our fourth generation model.

The dominance importance and ensemble importance are considered in an iterative fashion, as done in previous work of one of the authors.\cite{vanpoucke:PolymerInt2022,VerdingP:AdvMaterTech2025} An exhaustive set of sub-models are trained, with one, two or three features (from the full feature list of the third generation model) systematically disabled. This allows us to determine the influence of each individual feature on the model's performance. Five features are found to have a low importance ($<85$\%) on average over all models. Therefore, only the six features with an average ensemble importance $>90$\%  are kept for the fourth generation (\textit{cf.} SI Table VI).

\begin{enumerate}
    \item $\textit{T}$
    \item $[Cu] \times time$
    \item $[Cu]^3$
    \item $e^{[Cu]}$
    \item $\frac {[Cu]}{time}$
    \item $\ln ([Cu] \times time)$
\end{enumerate}

\begin{figure*}[!tb]
  \includegraphics[width=\linewidth]{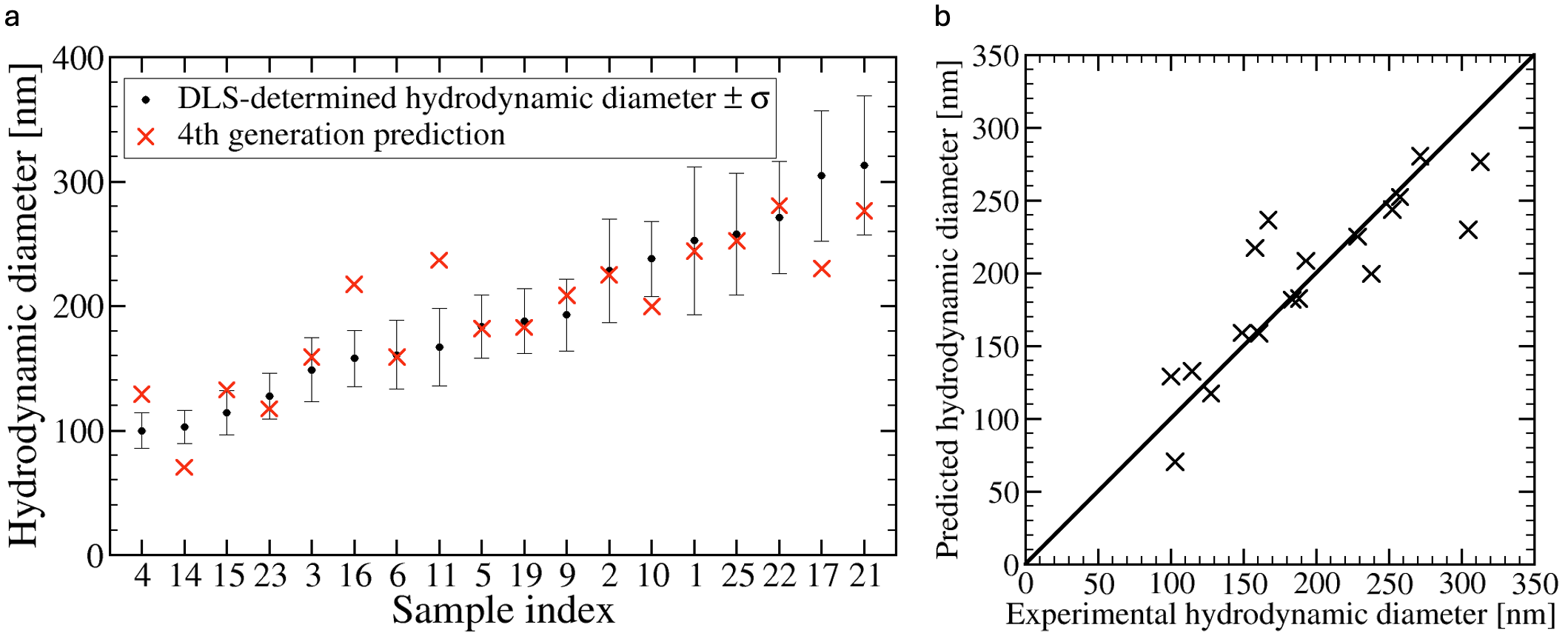}
  \caption{a) Plot of the experimentally measured hydrodynamic diameter and their standard deviation in ascending order (black) with the corresponding predicted hydrodynamic diameter of the fourth generation ensemble model (red); b) Parity plot of the actual, experimental hydrodynamic diameter against the corresponding predicted hydrodynamic diameter using the fourth generation ensemble model.}
  \label{Fig:4th_generation}
\end{figure*}

The predicted hydrodynamic diameters resulting from our fourth generation ensemble regression model are shown in figure~\ref{Fig:4th_generation}. In this model, each of the six features has an ensemble importance of $\geq 98.0$\% (LASSO regularization, $\alpha = 0.1$). The quality of the model is only very slightly below the $19$-feature model, showing a significant improvement with respect to overfitting (\textit{cf.} Table~\ref{table:model generations performance}). Because the size of the model has been greatly reduced, the MAE for the OOB predictions decreases from $52.18$ nm for the third generation to $40.92$ nm for the fourth generation, representing a marked improvement. It should be noted that this predictive performance is obtained exclusively for mono-modal DLS distributions. Consequently, the regression model is intended for synthesis conditions yielding a unique hydrodynamic diameter and should therefore be combined with the classification model presented below to identify its applicability domain.

The predictive performance obtained in this work should be interpreted in the context of the limited dataset size. Vanpoucke \textit{et al.}\cite{vanpoucke:JournalApplPhys2020}, demonstrate that model instances trained on the same dataset can yield substantially different performance metrics due to a pronounced “chance factor” inherent to small datasets. Furthermore, predictive performance improves systematically with increasing dataset size, following a power-law behavior, implying that direct comparison of performance metrics across studies with different dataset sizes is not straightforward. In this context, the $R^2$ value of $0.74$ obtained here is consistent with expectations for regression tasks based on fewer than 20–30 experimental data points, and should be interpreted as a good and meaningful predictive capability rather than an absolute measure of model quality. Importantly, ensemble-based approaches have been shown to improve the robustness and reliability of predictions in small-data regimes by reducing sensitivity to individual data points. This supports the use of the ensemble modeling strategy adopted in this work. When compared to literature reports, it is important to note that higher predictive accuracies are often obtained for classification tasks or simpler targets such as yield, as demonstrated by Guda \textit{et al.}\cite{guda:JournalPhysChemC2023}, and Williams \textit{et  al.}\cite{williams:AppliedNano2023}, whereas more complex regression tasks, such as predicting nanoparticle size or morphology, typically result in lower predictive performance, as observed by Pellegrino \textit{et al.}\cite{pellegrino:ScientificRep2020}. Therefore, the predictive performance obtained in this work is consistent with the expected accuracy for small-data, multi-parameter regression problems.

Since the average of the ensemble model is equivalent to a single model instance with averaged coefficients \cite{vanpoucke:JournalApplPhys2020}, our fourth generation model can be formulated as follows:

\begin{equation}
  \begin{aligned}
  \text{Hydrodynamic diameter}=\; -194.66-12.59T-36.16[Cu]t \\
   - 123.47[Cu]^3+108.12 e^{[Cu]}+\frac{32.30[Cu]}{t} + 63.00 \ln([Cu]t) \label{eq:4th generation_std}
  \end{aligned}
\end{equation}

In Equation~\ref{eq:4th generation_std}, all feature values are standardized prior to training, meaning the coefficients apply to standardized inputs (\textit{cf.} SI Table VII). When standardized experimental feature values are inserted into Equation~\ref{eq:4th generation_std}, the model directly returns the predicted hydrodynamic diameter in nm.

\begin{equation}
  \begin{aligned}
  \text{Hydrodynamic diameter}=\;-4.60\times10^4-1.56T \\
  -9.73\times10^2[Cu]t-5.64 \times 10^8[Cu]^3 + 4.66 \times 10^4e^{[Cu]} \\
  +\frac{3.49\times 10^4[Cu]}{t}+68.46\ln([Cu]t) \label{eq:4th generation}
  \end{aligned}
\end{equation}

Equation~\ref{eq:4th generation} gives the non-standardized form of the fourth generation model. Here, $[Cu]$ in molar, $time$ in minutes and $\textit{T}$ in $^\circ C$ can be used directly to obtain the hydrodynamic diameter in nm.

\subsubsection{Statistical approach (DoE)}

\begin{figure*}[!tb]
  \includegraphics[width=\textwidth]{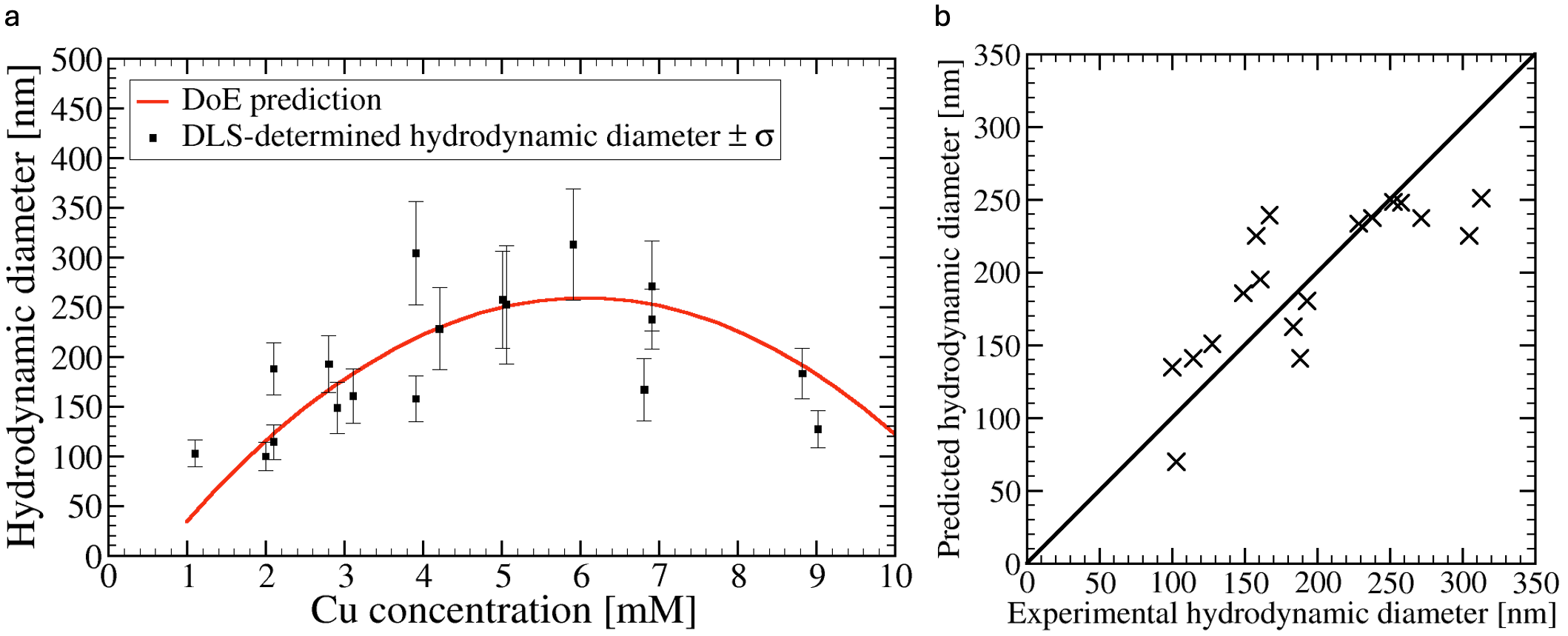}
  \caption{a) Plot of the experimental measured hydrodynamic diameters and their standard deviation (black) and the statistical model (red); b) Parity plot of the actual, experimental hydrodynamic diameter against the corresponding predicted hydrodynamic diameter using the statistical model.}
  \label{Fig:DoE model}
\end{figure*}

In the classical statistical approach, only the concentration of the Cu precursor appears to significantly influence the hydrodynamic diameter, resulting in a quadratic model shown in figure~\ref{Fig:DoE model} and given by Equation~\ref{eq:DoE model}. Here, [Cu] refers to the non-standardized precursor concentration in molar units, and the equation returns the hydrodynamic diameter directly in nm. Using equivalent out-of-sample validation procedures (LOOCV for the DoE model and OOB validation for the ensemble ML model), Table~\ref{table:model generations performance} shows that both approaches exhibit comparable generalization performance (MAE $= 40.46$ nm and $40.92$ nm, respectively). However, the final ensemble model achieves a lower prediction error (MAE $= 23.81$ nm) than the final DoE model (MAE $= 33.54$ nm), while simultaneously retaining all experimentally relevant synthesis parameters within the model, making it more suitable as a synthesis-guidance tool.

\begin{equation}
  \begin{aligned}
  \text{Hydrodynamic diameter}=&-63.24+106\,542.95[Cu]\\
  &-8\,801\,229.71[Cu]^2 \label{eq:DoE model}
  \end{aligned} 
\end{equation}

Because individual data points have a large impact in small datasets, models trained on different subsets capture slightly different aspects of the available data. By averaging the predictions of multiple such models, the influence of noise and outliers is reduced, resulting in a more accurate and robust final prediction.\cite{vanpoucke:JournalApplPhys2020, vanpoucke:PolymerInt2022} On the other hand, the classical regression model consists of only two terms containing the precursor concentration, compared to the six features in the ensemble regression model, which results in a simpler model and therefore lower computational cost. Comparison of the ML and DoE models indicates that Cu concentration is an important factor, as both contain it as a polynomial term. In addition, the influence of time and temperature are of relevance as well (as known from experiment) which is captured by the ML models only.

Since the DoE model only retained two features, we attempted to further reduce the fourth generation model by removal of additional features, based on dominance importance analyses (\textit{cf.} SI Table VIII and IX). First $\frac {[Cu]}{time}$ was removed, resulting in a slight performance reduction. $R^2$ dropped from $0.74$ to $0.72$ and the MAE increased from $23.81$ nm to $25.10$ nm. On the OOB data, both models have almost the same MAE, so despite the slightly lower performance there is no increase in generalization. Next, also $[Cu] \times time$ was removed, as it was assumed to be covered by the $\ln ([Cu] \times time)$ feature. Now the overall performance further dropped with an $R^2$ of $0.69$ and an MAE of $27.14$ nm. In this case, generalization does increase significantly with a training MAE of $26.26$ nm and an OOB MAE of $36.69$ nm. Finally, also $\ln ([Cu] \times time)$ was removed from the feature list, to arrive at a three feature model. Again, the generalization of the model increases with a training MAE of $26.56$ nm and an OOB MAE of $32.85$ nm, while maintaining similar final ensemble model performance with an $R^2$ of $0.69$ and an MAE of $26.96$ nm.

This demonstrates the delicate balance between model performance and overall model generalization. Reducing model size does give significant improvements in generalization, but comes with a cost in performance. While the fourth generation model with six features had an MAE of only $23.81$ nm, this increased to $26.96$ nm for the three feature model, getting closer to the $33.54$ nm MAE of the two feature DoE model.

\subsubsection{ML guided synthesis: model validation}

For three target hydrodynamic diameters ($115$ nm, $220$ nm and $275$ nm) Equation~\ref{eq:4th generation} is used to generate two candidate synthesis parameter sets (temperature, time, and concentration) per target hydrodynamic diameter (\textit{cf.} SI Table X). The target synthesis conditions were deliberately selected within the experimentally investigated synthesis parameter space to avoid extrapolation. The corresponding experiments are carried out using the same procedure as for the original dataset. For two individual cases (samples $27$ \& $28$), the obtained hydrodynamic diameters fall within the experimental uncertainty of the targeted values (Table~\ref{table:Synthesis prediction}). In contrast, samples $26$, $29$, $30$, and $31$ failed to yield hydrodynamic diameters that fell within the experimental uncertainty of their respective target values. For the largest target hydrodynamic diameter (samples $30$ \& $31$), neither of the proposed parameter sets yields the expected hydrodynamic diameter. It is interesting to note that in cases where the synthesis suggestions fail, the particles are always too small. This systematic underestimation is consistent with the restricted applicability domain of the regression model. Since the model was trained exclusively on mono-modal DLS distributions, larger hydrodynamic particle populations associated with the excluded multi-modal samples are underrepresented during model development. Consequently, predictions towards the upper end of the investigated hydrodynamic diameter range should be interpreted with appropriate caution, despite remaining within the investigated synthesis parameter space.

To provide a comparison, the same target hydrodynamic diameters can be evaluated using the DoE model (Equation~\ref{eq:DoE model}). However, due to the strongly reduced feature set in the DoE model, this approach only suggests Cu concentrations for the 115 nm and 220 nm targets, as no real solutions exist for the 275 nm target. In addition, the DoE model ignores the role of time and temperature in the synthesis process (\textit{cf.} SI Table XI). This highlights a key limitation of the DoE approach: while it can describe trends within the explored parameter space, it is less suitable for guiding synthesis towards specific target hydrodynamic diameters. In contrast, the ML model enables a more flexible exploration of synthesis conditions and can propose multiple parameter combinations to approach a desired target hydrodynamic diameter.

\begin{table}[!tb]
  \caption{Comparison of targeted hydrodynamic diameters, experimentally obtained hydrodynamic diameters and their $\sigma$ from predicted syntheses routes with the fourth generation ensemble model.}
  \label{table:Synthesis prediction}
  \centering
  \begin{tabular}{cccc}
    \hline
    Sample & Targeted $[nm]$ & Experimental $[nm]$ \\
    \hline
    $26$ & $115$ & $70.78~(\pm 4.68$) \\
    $27$ & $115$ & $117.90~(\pm 32.06$) \\
    $28$ & $220$ & $235.60~(\pm 34.57$) \\
    $29$ & $220$ & $128.60~(\pm 49.54$) \\
    $30$ & $275$ & $165.10~(\pm 41.04$) \\
    $31$ & $275$ & $189.60~(\pm 59.50$) \\
    \hline
  \end{tabular}
\end{table}

This validation is limited to only three target hydrodynamic diameters and therefore does not allow for a statistically robust assessment of predictive accuracy. Of the six proposed synthesis conditions (two for each target hydrodynamic diameter), two yielded hydrodynamic diameters within the experimental uncertainty of their respective target values. These successful syntheses correspond to two of the three investigated target hydrodynamic diameters. In contrast, the DoE model only predicts the required Cu concentration and therefore still requires manual trial-and-error optimization of temperature and time, increasing the number of required experiments. The limited number of experimental iterations required to reach the target hydrodynamic diameters suggests that ML guided synthesis can reduce experimental effort, and accordingly reduce resource consumption. In addition, the use of a simple, interpretable model expressed as an explicit analytical equation further limits computational overhead during model evaluation.

\subsection{DLS-derived hydrodynamic diameter classification}

\begin{table}[!t]
  \caption{Performance metrics of binary classification models.}
  \label{table:Classification performance}
  \centering
  \begin{tabular}{lcc}
    \hline
         & Accuracy & $\kappa$ \\
    \hline
    GPT-J &  $0.48$ & $-0.04$ \\
    LLAMA 3.1 & $0.64$ & $0.28$ \\
    Random forest & $0.56$ & $0.11$ \\
    \hline
  \end{tabular}
\end{table}

In addition to the regression model, also a binary classification model is created. The goal of this model is to determine whether the obtained Cu NPs will be \textit{large} (`1') or \textit{small} (`0'), based on the synthesis parameters.

We compare complex LLMs, fine-tuned via transfer learning, with a classical random forest model trained from scratch. Initial LLM training runs limited to $25$ epochs did not yield satisfactory performance, as the models failed to learn the prompt-completion structure. Consequently, invalid outputs were generated for the test data (\textit{e.g.}, numerical artifacts such as $-9223372036854775808$) instead of valid binary labels (`0' or `1'), preventing the calculation of meaningful performance metrics. With a higher number of epochs, however, the fine-tuned model consistently generated valid predictions (`0' or `1'). (\textit{cf.} SI Figure 10)

GPT-J achieved an accuracy of $0.48$ and a $\kappa=-0.04$, LLAMA 3.1 achieved an accuracy of $0.64$ and $\kappa=0.28$, and the random forest classifier achieved an accuracy of $0.56$ and a $\kappa=0.11$ (Table~\ref{table:Classification performance}). For all three models, the overall performance is relatively low. One of the possible reasons for this limited performance is the very limited size of the data set. Small data sets already pose major challenges for classic ML, such as regression, but the use of such small data sets is particularly challenging for complex models as LLMs.

Overall, the results do not provide evidence that fine-tuned LLMs offer a practical advantage over a classical ML approach for this small-data application. Since all evaluated models exhibited only limited predictive performance, no further investigation of this classification was pursued in this work. It is clear that these require a much larger data set to achieve proper performance.\cite{van-herck:ChemScience2024} Considering that LLMs are much more complex than classical models makes them less efficient with regard to the computational cost and thus energy requirements.

\subsection{DLS distribution modality classification}

\begin{figure}[!tb]
  \includegraphics[width=\columnwidth]{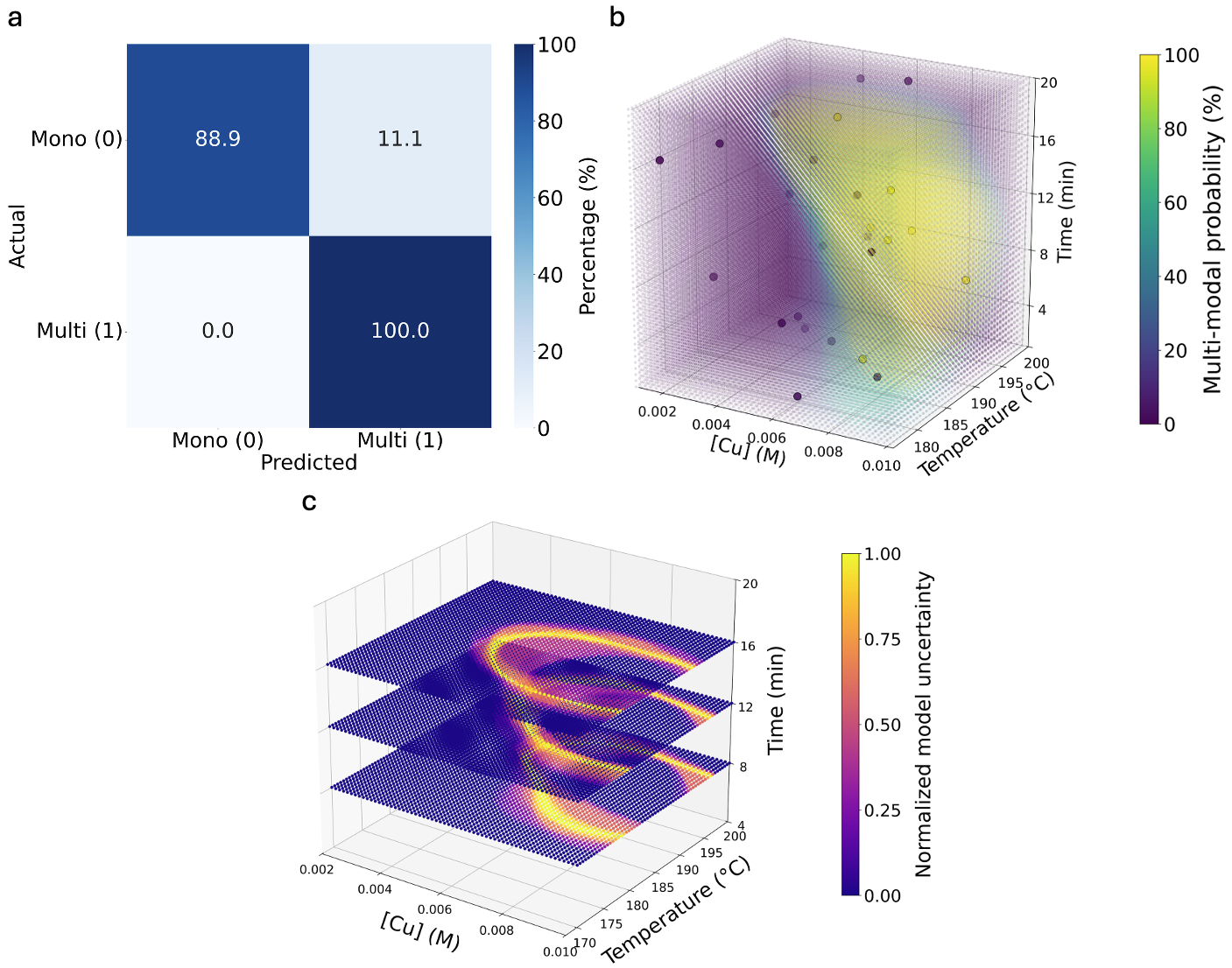}
  \caption{a) Confusion matrix of the SVM classifier for mono- and multi-modal DLS distributions; b) Visualization of the SVM ensemble predictions in the synthesis parameter space. The color scale represents the predicted probability of a multi-modal DLS distribution; c) 3D uncertainty map showing how uncertainty is concentrated around the models decision boundary, while large portions of the parameter space show almost zero uncertainty.}
  \label{Fig:results svm}
\end{figure}

As previously discussed, the predictive performance of the regression model strongly depends on restricting the dataset to mono-modal DLS hydrodynamic diameter distributions. However, the occurrence of mono- and multi-modal distributions cannot be determined \textit{a priori} from the synthesis parameters. Therefore, a classification model is introduced to predict the modality of the DLS distribution based on the synthesis conditions.

The model achieves an accuracy of $0.92$ and a $\kappa=0.82$, indicating strong predictive performance despite the limited dataset size. (\textit{cf.} Figure~\ref{Fig:results svm}a) These results demonstrate that the synthesis parameters contain sufficient information to distinguish between mono- and multi-modal outcomes. The distribution of predicted outcomes across the synthesis parameter space is visualized in figure~\ref{Fig:results svm}b. Model stability was subsequently evaluated by calculating the variance in predictions across 100 independently trained SVM models. The resulting uncertainty map (figure~\ref{Fig:results svm}c) demonstrates that the uncertainty is not uniformly distributed throughout the feature space, but instead remains concentrated within a relatively narrow transition region surrounding the decision boundary. Importantly, this classification task provides a necessary complement to the regression model. As previously shown, the regression model is applicable only to mono-modal DLS distributions, and is not applicable to multi-modal distributions. The SVM model can therefore be interpreted as a screening step, enabling the identification of conditions under which the regression model is expected to yield reliable predictions.

This classification task provides additional insight into the conditions under which the regression model is expected to perform reliably. By identifying whether a given set of synthesis parameters is likely to yield a mono- or multi-modal distribution, the model helps to define the domain in which quantitative hydrodynamic diameter predictions are valid. This highlights the potential of combining classification and regression approaches to improve the applicability of modeling strategies in small-data experimental settings.

\section{Conclusions}\label{sec:conclusion}

Despite the small size of the data set, it is possible to predict the hydrodynamic diameter of Cu NPs quantitatively based on synthesis parameters using classical ensemble regression ML, achieving an $R^2$ of $0.74$, an MAE of $23.81$ nm, and an RMSE of $33.03$ nm. For most data points, the predicted hydrodynamic diameter falls within the experimental standard deviation. We were able to reduce the number of features in our final model to six, which improved generalization and reduced model size. Using equivalent out-of-sample validation procedures, the ensemble ML model (OOB: MAE $= 40.92$ nm, RMSE $= 49.90$ nm) and the DoE model (LOOCV: MAE $= 40.46$ nm, RMSE $= 48.22$ nm) showed comparable generalization. The final ensemble ML model nevertheless retained a slightly lower prediction error (MAE $= 23.81$ nm) than the final DoE model (MAE $= 33.54$ nm) while simultaneously incorporating all experimentally relevant synthesis parameters, making it more suitable for synthesis guidance across the investigated parameter space. As with all models developed in this work, the proposed workflow is intended for interpolation within the experimentally investigated synthesis parameter space. Extending this parameter space through additional experiments therefore represents a possible direction for future model development.

By in-house constructing a consistent data set, we overcome limitations associated with biased or incomplete literature data. The use of interpretable models allowed us to identify the synthesis parameters with the strongest influence on particle properties, providing fundamental insights alongside predictive power. As validation, the final model is used to propose two synthesis routes for each of three target hydrodynamic diameters. Of the six proposed synthesis conditions, two yielded hydrodynamic diameters within the experimental uncertainty of their respective target values, corresponding to two of the three investigated target hydrodynamic diameters. Although limited in scope, these validation results indicate that ML can provide useful guidance for experimental synthesis even with a minimal dataset. These results further show that the synthesis outcome is not only characterized by hydrodynamic diameter, but also by the modality of the resulting DLS-derived hydrodynamic diameter distribution. Consequently, quantitative hydrodynamic diameter prediction is restricted to mono-modal DLS distributions. The accompanying SVM classification model can be used as a screening step to identify synthesis conditions for which the regression model is expected to provide reliable quantitative predictions.

We additionally explored the applicability of LLMs for synthesis prediction in this small-data regime. While LLMs offer conceptual advantages, such as operating directly on recipe-style inputs without explicit feature engineering, our results indicate that they fail to identify reliable relationships between synthesis parameters and synthesis outcome from limited experimental data. While larger data sets might improve LLM performance, generating such data would conflict with the objective of data-efficient synthesis optimization in this current lab-scale experimental setting. These findings suggest that, for lab-scale NP synthesis with data sets typical in experimental materials research, carefully curated small data combined with interpretable classical ML models provides a more reliable approach than deploying more complex LLMs. In addition, the results indicate that ML guided synthesis has the potential to improve the efficiency of experimental optimization. The limited number of experimental iterations required to reach the target hydrodynamic diameters suggests that this approach can reduce experimental effort and, accordingly, resource consumption.

Although this work focused on Cu NPs as a model system, the approach is broadly applicable to other materials. Future work could integrate such ML workflows into closed-loop experimentation and explore larger, more diverse data sets. Overall, this strategy contributes to more efficient, interpretable, and sustainable materials development.

\section*{Author contributions}

Brent Motmans: Data curation, Formal analysis, Investigation, Software, Writing–original draft, Writing–review \& editing. Digvijay Ghogare: Supervision, Writing–review \& editing. Thijs G. I. van Wijk: Supervision, Writing–review \& editing. Joren Van Herck: Formal analysis, Software, Writing–review \& editing. Saba Heidarian: Formal analysis, Software, Writing–review \& editing. Pieter De Meyer: Formal analysis, Software, Writing–review \& editing. Berend Smit: Supervision, Writing–review \& editing. An Hardy: Conceptualization, Supervision, Writing–review \& editing. Danny E. P. Vanpoucke: Conceptualization, Supervision, Writing–review \& editing.

\section*{Conflicts of interest}

There are no conflicts to declare.

\section*{Data availability}

All data supporting this article are included in the manuscript and Supplementary Information.

\section*{Acknowledgements}

The authors acknowledge Sander Stulens and Romy Van Geijn for performing the selected area electron diffraction (SAED) and TEM measurements. This work was financially supported by the Special Research Fund (BOF) via the BOF project, Belgium R15949. The computational resources and services used in this work were provided by the VSC (Flemish Supercomputer Center), funded by the Research Foundation Flanders (FWO) and the Flemish Government--department WEWIS.

\section*{Supporting information}

The following files are available free of charge.
\begin{itemize}
  \item Supplementary\_information.pdf: Experimental synthesis parameters and datasets; DLS, UV-Vis, and SAED characterization data; ML model development and validation; DoE, SVM, and LLM analyses.
\end{itemize}
\bibliography{CuNP}

\clearpage
\onecolumngrid

\setcounter{figure}{0}
\setcounter{table}{0}

\section*{Supporting information for "Predictive Inorganic Synthesis based on Machine Learning using Small Data sets: a case study of Hydrodynamic Diameter-controlled Copper Nanoparticles"}
\begin{table}[!h]
\caption{Sampled synthesis parameters used for the 25 experimental syntheses.
(a) Theoretical concentration of Cu(OAc)$_2$ precursor in 5 ml ethylene glycol in mM;
(b) Effective Cu concentration of Cu(OAc)$_2$ used;
(c) Volume of 2.75 M TMAH in H$_2$O added to 5 ml ethylene glycol in $\mu$l;
(d) Time in minutes reaction mixture is kept on defined temperature;
(e) Temperature in $^\circ$C of reaction mixture for defined reaction time.}
\label{tab:S1}
\begin{tabular}{lccccc}
\hline
Sample & Theoretical Cu conc.$^{a}$ [mM] & Effective Cu prec.$^{b}$ [mM] & TMAH$^{c}$ [$\mu$l] & $Time^{d}$ [min] & Temp.$^{e}$ [$^\circ$C] \\
\hline
1  & 5.00 & 5.06 & 36.0 &  8 & 196 \\
2  & 4.00 & 4.21 & 29.0 &  4 & 188 \\
3  & 3.00 & 2.91 & 21.8 &  6 & 198 \\
4  & 2.00 & 2.00 & 14.6 &  7 & 184 \\
5  & 9.00 & 8.82 & 65.4 &  4 & 180 \\
6  & 3.00 & 3.11 & 21.8 &  1 & 194 \\
7  &10.00 &10.22 & 72.8 &  9 & 188 \\
8  & 9.00 & 8.82 & 65.4 & 16 & 182 \\
9  & 3.00 & 2.80 & 21.8 & 17 & 182 \\
10 & 7.00 & 6.91 & 51.0 & 20 & 194 \\
11 & 7.00 & 6.81 & 51.0 & 14 & 186 \\
12 & 6.00 & 6.01 & 43.6 & 10 & 192 \\
13 & 8.00 & 8.31 & 58.2 & 12 & 184 \\
14 & 1.00 & 1.10 &  7.2 & 15 & 198 \\
15 & 2.00 & 2.10 & 14.6 & 10 & 196 \\
16 & 4.00 & 3.91 & 29.0 & 18 & 200 \\
17 & 4.00 & 3.91 & 29.0 & 14 & 192 \\
18 & 8.00 & 8.31 & 58.2 &  5 & 180 \\
19 & 2.00 & 2.10 & 14.6 & 17 & 176 \\
20 & 8.00 & 7.91 & 58.2 & 11 & 190 \\
21 & 6.00 & 5.91 & 43.6 &  7 & 178 \\
22 & 7.00 & 6.91 & 51.0 &  3 & 176 \\
23 & 9.00 & 9.02 & 65.4 & 13 & 178 \\
24 & 6.00 & 6.11 & 43.6 & 19 & 186 \\
25 & 5.00 & 5.01 & 36.0 &  2 & 190 \\
\hline
\end{tabular}
\end{table}

\begin{table}[!h]
\caption{ Data set A containing 25 datapoints with particle size, the smallest size detected by DLS.
(a) Effective concentration of Cu(OAc)$_2$ precursor in 5 ml ethylene glycol in mM;
(b) Volume of 2.75 M TMAH in H$_2$O added to 5 ml ethylene glycol in $\mu$l;
(c) Time in minutes reaction mixture is kept on defined temperature;
(d) Temperature in $^\circ$C of reaction mixture for defined reaction time.
(e) Hydrodynamic diameter of the NPs and the standard deviation ($\sigma$) in nm as measured using DLS.}
\label{tab:S1}
\begin{tabular}{cccccc}
\hline
Sample & Effective Cu conc.$^{a}$ [mM] & TMAH$^{b}$ [$\mu$l] & $Time^{c}$ [min] & Temp.$^{d}$ [$^\circ$C] & Size$^{e}$ [nm]\\
\hline
1 & 5.06 & 36.0 & 8 & 196 & 252.50 ($\pm$59.37) \\
2 & 4.21 & 29.0 & 4 & 188 & 228.40 ($\pm$41.47) \\
3 & 2.91 & 21.8 & 6 & 198 & 148.80 ($\pm$25.56) \\
4 & 2.00 & 14.6 & 7 & 184 & 99.90 ($\pm$14.07) \\
5 & 8.82 & 65.4 & 4 & 180 & 183.30 ($\pm$25.42) \\
6 & 3.11 & 21.8 & 1 & 194 & 160.70 ($\pm$27.62) \\
7 &10.22 & 72.8 & 9 & 188 & 64.65 ($\pm$8.80) \\
8 & 8.82 & 65.4 & 16 & 182 & 40.85 ($\pm$5.20) \\
9 & 2.80 & 21.8 & 17 & 182 & 192.90 ($\pm$28.88) \\
10 & 6.91 & 51.0 & 20 & 194 & 237.90 ($\pm$30.29) \\
11 & 6.81 & 51.0 & 14 & 186 & 167.00 ($\pm$31.15) \\
12 & 6.01 & 43.6 & 10 & 192 & 60.10 ($\pm$6.88) \\
13 & 8.31 & 58.2 & 12 & 184 & 66.02 ($\pm$11.94) \\
14 & 1.10 &  7.2 & 15 & 198 & 102.90 ($\pm$13.51) \\
15 & 2.10 & 14.6 & 10 & 196 & 114.40 ($\pm$17.64) \\
16 & 3.91 & 29.0 & 18 & 200 & 157.90 ($\pm$22.70) \\
17 & 3.91 & 29.0 & 14 & 192 & 304.50 ($\pm$52.21) \\
18 & 8.31 & 58.2 & 5 & 180 & 59.09 ($\pm$6.13) \\
19 & 2.10 & 14.6 & 17 & 176 & 188.10 ($\pm$26.04) \\
20 & 7.91 & 58.2 & 11 & 190 & 50.36 ($\pm$8.16) \\
21 & 5.91 & 43.6 & 7 & 178 & 312.90 ($\pm$55.82) \\
22 & 6.91 & 51.0 & 3 & 176 & 271.60 ($\pm$45.13) \\
23 & 9.02 & 65.4 & 13 & 178 & 127.60 ($\pm$18.55) \\
24 & 6.11 & 43.6 & 19 & 186 & 58.69 ($\pm$10.61) \\
25 & 5.01 & 36.0 & 2 & 190 & 257.60 ($\pm$48.85) \\
\hline
\end{tabular}
\end{table}

\begin{table}[!h]
\caption{ Data set B containing 25 datapoints with particle size, the size with the highest percentage contribution in DLS.
(a) Effective Cu concentration of Cu(OAc)$_2$ used in 5 ml ethylene glycol;
(b) Volume of 2.75 M TMAH in H$_2$O added to 5 ml ethylene glycol in $\mu$l;
(c) Time in minutes reaction mixture is kept on defined temperature;
(d) Temperature in $^\circ$C of reaction mixture for defined reaction time;
(e) Hydrodynamic diameter of the NPs and the $\sigma$ in nm as measured using DLS.}
\label{tab:S3}
\begin{tabular}{cccccc}
\hline
Sample & Effective Cu conc.$^{a}$ [mM] & TMAH$^{b}$ [$\mu$l] & $Time^{c}$ [min] & Temp.$^{d}$ [$^\circ$C] & Size$^{e}$ [nm] \\
\hline
1 & 5.06 & 36.0 & 8 & 196 & 252.50 ($\pm$59.37) \\
2 & 4.21 & 29.0 & 4 & 188 & 228.40 ($\pm$41.47) \\
3 & 2.91 & 21.8 & 6 & 198 & 148.80 ($\pm$25.56) \\
4 & 2.00 & 14.6 & 7 & 184 & 99.90 ($\pm$14.07) \\
5 & 8.82 & 65.4 & 4 & 180 & 183.30 ($\pm$25.42) \\
6 & 3.11 & 21.8 & 1 & 194 & 160.70 ($\pm$27.62) \\
7 & 10.22 & 72.8 & 9 & 188 & 204.00 ($\pm$34.54) \\
8 & 8.82 & 65.4 & 16 & 182 & 222.30 ($\pm$31.62) \\
9 & 2.80 & 21.8 & 17 & 182 & 192.90 ($\pm$28.88) \\
10 & 6.91 & 51.0 & 20 & 194 & 237.90 ($\pm$30.29) \\
11 & 6.81 & 51.0 & 14 & 186 & 167.00 ($\pm$31.15) \\
12 & 6.01 & 43.6 & 10 & 192 & 269.50 ($\pm$60.33) \\
13 & 8.31 & 58.2 & 12 & 184 & 66.02 ($\pm$11.94) \\
14 & 1.10 & 7.2 & 15 & 198 & 102.90 ($\pm$13.51) \\
15 & 2.10 & 14.6 & 10 & 196 & 114.40 ($\pm$17.64) \\
16 & 3.91 & 29.0 & 18 & 200 & 157.90 ($\pm$22.70) \\
17 & 3.91 & 29.0 & 14 & 192 & 304.50 ($\pm$52.21) \\
18 & 8.31 & 58.2 & 5 & 180 & 184.90 ($\pm$26.97) \\
19 & 2.10 & 14.6 & 17 & 176 & 188.10 ($\pm$26.04) \\
20 & 7.91 & 58.2 & 11 & 190 & 50.36 ($\pm$8.16) \\
21 & 5.91 & 43.6 & 7 & 178 & 312.90 ($\pm$55.82) \\
22 & 6.91 & 51.0 & 3 & 176 & 271.60 ($\pm$45.13) \\
23 & 9.02 & 65.4 & 13 & 178 & 127.60 ($\pm$18.55) \\
24 & 6.11 & 43.6 & 19 & 186 & 58.69 ($\pm$10.61) \\
25 & 5.01 & 36.0 & 2 & 190 & 257.60 ($\pm$48.85) \\
\hline
\end{tabular}
\end{table}

\begin{table}[!h]
\centering
\caption{Data set C containing 18 datapoints with monomodal DLS particle-size distribution.
(a) Effective Cu concentration of Cu(OAc)$_2$ used in 5 ml ethylene glycol in mM;
(b) Volume of 2.75 M TMAH in H$_2$O added to 5 ml ethylene glycol in $\mu$l;
(c) Time in minutes reaction mixture is kept on defined temperature;
(d) Temperature in $^\circ$C of reaction mixture for defined reaction time;
(e) Hydrodynamic diameter of the NPs and the $\sigma$ in nm as measured using DLS.}
\label{tab:S4}
\begin{tabular}{cccccc}
\hline
Sample & Effective Cu conc.$^{a}$ [mM] & TMAH$^{b}$ [$\mu$l] & $Time^{c}$ [min] & Temp.$^{d}$ [$^\circ$C] & Size$^{e}$ [nm] \\
\hline
1 & 5.06 & 36.0 & 8 & 196 & 252.50 ($\pm$59.37) \\
2 & 4.21 & 29.0 & 4 & 188 & 228.40 ($\pm$41.47) \\
3 & 2.91 & 21.8 & 6 & 198 & 148.80 ($\pm$25.56) \\
4 & 2.00 & 14.6 & 7 & 184 & 99.90 ($\pm$14.07) \\
5 & 8.82 & 65.4 & 4 & 180 & 183.30 ($\pm$25.42) \\
6 & 3.11 & 21.8 & 1 & 194 & 160.70 ($\pm$27.62) \\
9 & 2.80 & 21.8 & 17 & 182 & 192.90 ($\pm$28.88) \\
10 & 6.91 & 51.0 & 20 & 194 & 237.90 ($\pm$30.29) \\
11 & 6.81 & 51.0 & 14 & 186 & 167.00 ($\pm$31.15) \\
14 & 1.10 & 7.2 & 15 & 198 & 102.90 ($\pm$13.51) \\
15 & 2.10 & 14.6 & 10 & 196 & 114.40 ($\pm$17.64) \\
16 & 3.91 & 29.0 & 18 & 200 & 157.90 ($\pm$22.70) \\
17 & 3.91 & 29.0 & 14 & 192 & 304.50 ($\pm$52.21) \\
19 & 2.10 & 14.6 & 17 & 176 & 188.10 ($\pm$26.04) \\
21 & 5.91 & 43.6 & 7 & 178 & 312.90 ($\pm$55.82) \\
22 & 6.91 & 51.0 & 3 & 176 & 271.60 ($\pm$45.13) \\
23 & 9.02 & 65.4 & 13 & 178 & 127.60 ($\pm$18.55) \\
25 & 5.01 & 36.0 & 2 & 190 & 257.60 ($\pm$48.85) \\
\hline
\end{tabular}
\end{table}

\begin{sidewaystable}
\centering
\caption{ Resulting sizes and SPR peaks of the different synthesis samples.
(a) Hydrodynamic diameter in nm of the Cu NPs and the $\sigma$ measured using DLS in ethanol;
(b) Percentage contribution of the respective peak to the DLS particle-size distribution;
(c) Due to their irrelevant size and minimal contribution to the DLS particle-size distribution, these peaks will not be considered during the further study;
(d) Location of the SPR peak in nm as measured in ethylene glycol using UV-Vis.}
\label{tab:S5}
\begin{tabular}{ccccccccc}
\hline
Sample & Size\_1$^{a}$ [nm] & Percentage\_1$^{b}$ [\%] & Size\_2$^{a}$ [nm] & Percentage\_2$^{b}$ [\%] & Size\_3$^{a}$ [nm] & Percentage\_3$^{b}$ [\%] & SPR$^{d}$ [nm] \\
\hline
1 & 252.50 ($\pm$59.37) & 99.9 & 5414.00 ($\pm$647.80)$^{c}$ & 0.1 & - & - & 613 \\
2 & 228.40 ($\pm$41.47) & 99.9 & 5135.00 ($\pm$775.60)$^{c}$ & 0.1 & - & - & 625 \\
3 & 148.80 ($\pm$25.56) & 100 & - & - & - & - & 612 \\
4 &  99.90 ($\pm$14.07) & 100 & - & - & - & - & 699 \\
5 & 183.30 ($\pm$25.42) & 100 & - & - & - & - & 597 \\
6 & 160.70 ($\pm$27.62) & 100 & - & - & - & - & 699 \\
7 & 64.65 ($\pm$8.80) & 29.1 & 204.00 ($\pm$34.54) & 70.9 & - & - & 592 \\
8 & 40.85 ($\pm$5.20) & 33.1 & 222.30 ($\pm$31.62) & 66.9 & - & - & 590 \\
9 & 192.90 ($\pm$28.88) & 100 & - & - & - & - & 609 \\
10 & 237.90 ($\pm$30.29) & 100 & - & - & - & - & 591 \\
11 & 167.00 ($\pm$31.15) & 98.9 & 5164.00 ($\pm$744.50)$^{c}$ & 1.1 & - & - & 636 \\
12 & 60.10 ($\pm$6.88) & 27.1 & 123.70 ($\pm$22.78) & 14.1 & 269.50 ($\pm$60.33) & 58.8 & 622 \\
13 & 66.02 ($\pm$11.94) & 53.0 & 126.20 ($\pm$18.36) & 4.7 & 259.90 ($\pm$45.83) & 42.3 & 611 \\
14 & 102.90 ($\pm$13.51) & 100 & - & - & - & - & 618 \\
15 & 114.40 ($\pm$17.64) & 100 & - & - & - & - & 619 \\
16 & 157.90 ($\pm$22.70) & 100 & - & - & - & - & 593 \\
17 & 304.50 ($\pm$52.21) & 100 & - & - & - & - & 616 \\
18 &  59.09 ($\pm$6.13) & 30.4 & 184.90 ($\pm$26.97) & 69.6 & - & - & 605 \\
19 & 188.10 ($\pm$26.04) & 99.7 & 5532.00 ($\pm$608.70)$^{c}$ & 0.3 & - & - & 626 \\
20 &  50.36 ($\pm$8.16) & 63.5 & 233.10 ($\pm$37.46) & 36.5 & - & - & 595 \\
21 & 312.90 ($\pm$55.82) & 100 & - & - & - & - & 609 \\
22 & 271.60 ($\pm$45.13) & 99.9 & 5380.00 ($\pm$654.70)$^{c}$ & 0.1 & - & - & 601 \\
23 & 127.60 ($\pm$18.55) & 100 & - & - & - & - & 595 \\
24 &  58.69 ($\pm$10.61) & 56.4 & 247.00 ($\pm$39.81) & 43.6 & - & - & 600 \\
25 & 257.60 ($\pm$48.85) & 100 & - & - & - & - & 597 \\
\hline
\end{tabular}
\end{sidewaystable}

\begin{figure}[h]
    \includegraphics[width=0.75\textwidth]{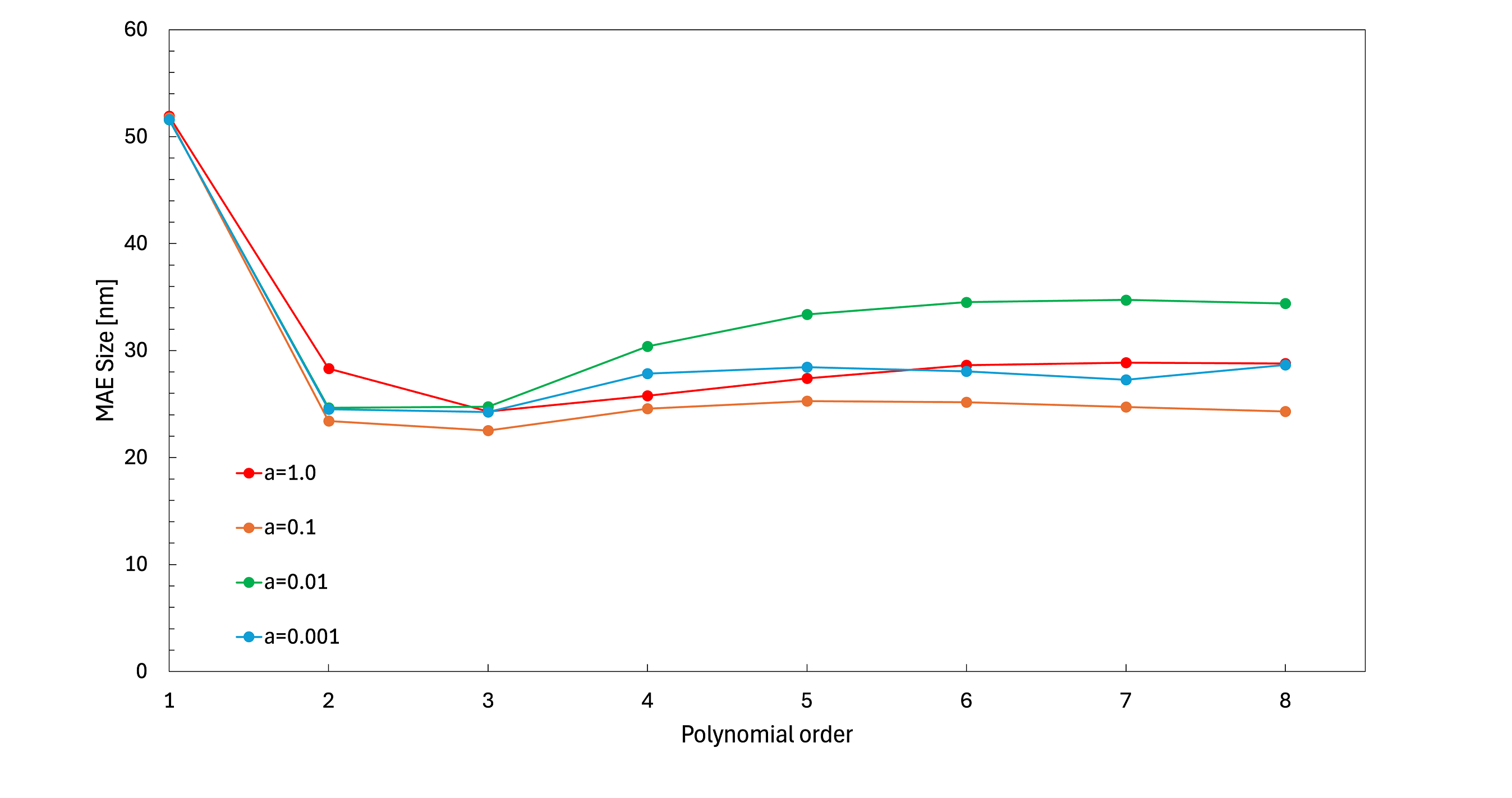}
    \caption{ Hyperparameter optimization for size prediction with data set C. LASSO regularized polynomials up to order eight are compared, and MAE of the final ensemble models is shown for $\alpha$=1.0 (red), $\alpha$=0.1 (orange), $\alpha$=0.01 (green), and $\alpha$=0.001 (blue). \label{Fig:Hyperparameter optimization}}
\end{figure}

\begin{figure}[h]
    \includegraphics[width=0.75\textwidth]{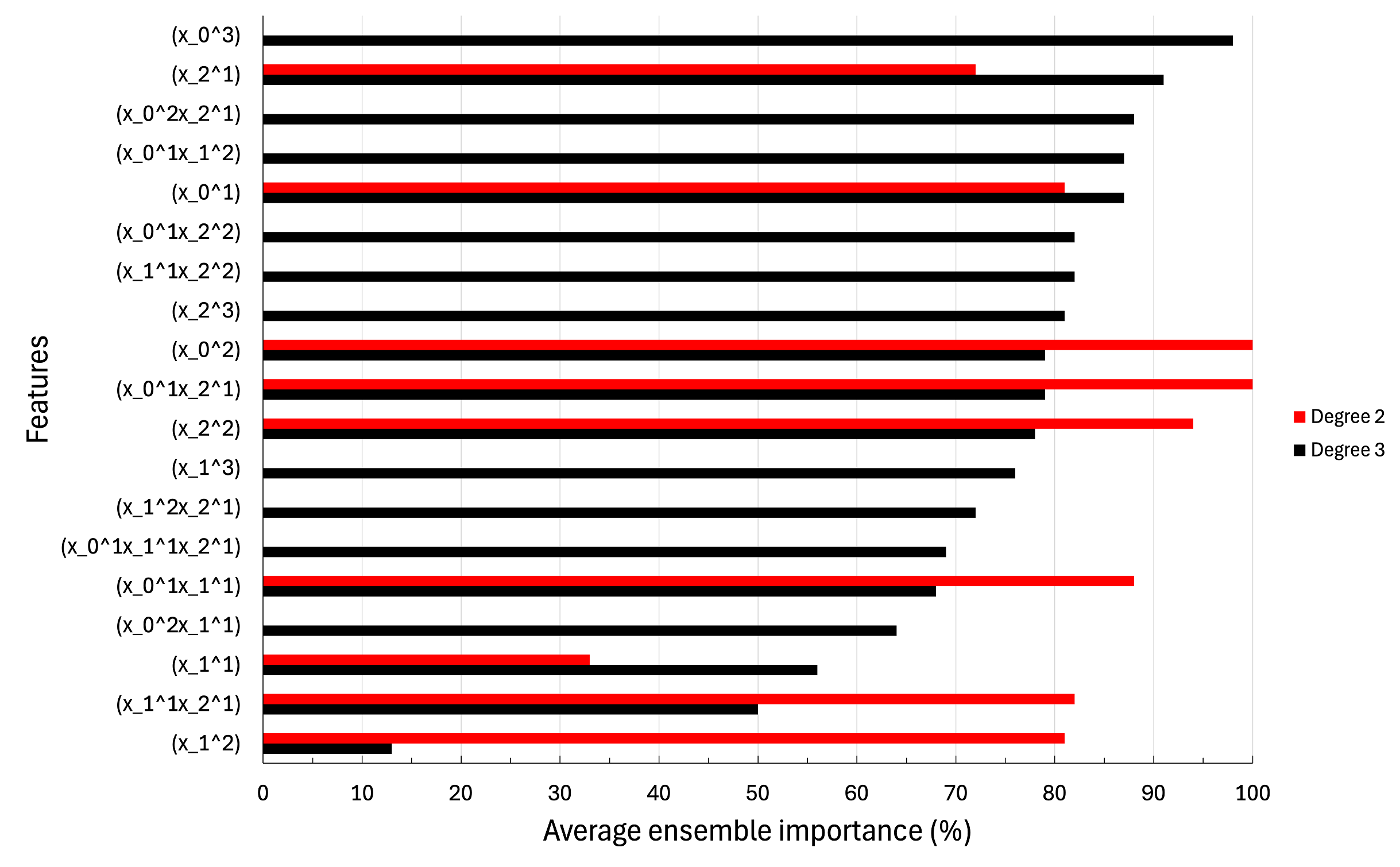}
    \caption{ Overview of feature importance, sorted by increasing importance in the third degree polynomial, in the polynomial models with degree two (red) and degree three (black), and trained on data set C with $\alpha$=0.1. x\_0 = Cu precursor concentration, x\_1 = Reaction temperature and x\_2 = Reaction time.\label{Fig:Ensemble Importance}}
\end{figure}

\begin{table}[h]

\caption{ Summary of the dominance importance analysis performed on the third generation ensemble model.}
\label{tab:S1}
\begin{tabular}{c|ccc}
\hline
Feature & Average EI [\%] & Minimum EI [\%] & Maximum EI [\%] \\
\hline
$[Cu]$  & 84.5 & 10.0 & 100 \\
\textit{T}  & 99.6 & 97.0 & 100 \\
time  & 82.2 & 54.0 & 100 \\
$[Cu]^2$  & 82.3 & 53.0 & 100 \\
$[Cu] \times T$  & 80.0 & 43.0 & 100 \\
$[Cu] \times time$  & 99.1 & 95.0 & 100 \\
$[Cu]^3$  & 98.0 & 85.0 & 100 \\
$e^{[Cu]}$  & 91.7 & 62.0 & 100 \\
$\frac{[Cu]}{time}$  & 95.1 & 84.0 & 100 \\
$\ln ([Cu] \times time)$  & 91.9 & 80.0 & 100 \\
$\frac{[time]}{\textit{T}}$  & 84.7 & 63.0 & 100 \\
\hline
\end{tabular}
\end{table}

\begin{table}[tb]

\caption{ Mean and standard deviation ($\sigma$) used to standardize the different features in the 4th generation ensemble model.}
\label{tab:S1}
\begin{tabular}{c|c@{\hspace{1cm}}c}
\hline
Feature & $\bar{x}$ & $\sigma$ \\
\hline
\textit{T} & $1.88 \times 10^2$ & $8.07 \times 10^0$ \\
$[Cu] \times time$ & $4.42 \times 10^{-2}$ & $3.72 \times 10^{-2}$ \\
$[Cu]^3$ & $1.75 \times 10^{-7}$ & $2.19 \times 10^{-7}$ \\
$e^{[Cu]}$ & $1.00 \times 10^0$ & $2.32 \times 10^{-3}$ \\
$\frac{[Cu]}{time}$ & $8.90 \times 10^{-4}$ & $9.27 \times 10^{-4}$ \\
$\ln ([Cu] \times time)$  & $-3.49 \times 10^0$ & $9.20 \times 10^1$ \\
\hline
\end{tabular}
\end{table}

\begin{table}[tb]

\caption{ Summary of the three best performing models resulting from the dominance importance analysis performed on the fourth generation ensemble model. 1 indicating the feature is included in the model, and 0 indicating the feature is not included.}
\label{tab:S1}
\begin{tabular}{c|c|c|c|c|c|cccc}
\hline
\textit{T} & $[Cu] \times time$ & $[Cu]^3$ & $e^{[Cu]}$ & $\frac{[Cu]}{time}$ & $\ln ([Cu] \times time)$ & MAE train & MAE OOB & RMSE train & RMSE OOB \\
\hline
1 & 1 & 1 & 1 & 0 & 1 & 24.45 & 37.69 & 32.47 & 45.57 \\
1 & 0 & 1 & 1 & 0 & 1 & 26.21 & 36.86 & 34.73 & 44.36 \\
1 & 0 & 1 & 1 & 0 & 0 & 26.66 & 34.43 & 35.33 & 42.58 \\
\hline
\end{tabular}
\end{table}

\begin{table}[tb]

\caption{ Because the engineered $e^{[Cu]}$ feature exhibits only limited variation over the investigated concentration range, dominance importance analysis is performed to assess whether its removal affects the predictive performance of the ensemble model. 1 indicating the feature is included in the model, and 0 indicating the feature is not included. None of the models excluding $e^{[Cu]}$ outperformed the original fourth generation ensemble model.}
\label{tab:S1}
\begin{tabular}{c|c|c|c|c|c|cccc}
\hline
\textit{T} & $[Cu] \times time$ & $[Cu]^3$ & $e^{[Cu]}$ & $\frac{[Cu]}{time}$ & $\ln ([Cu] \times time)$ & MAE train & MAE OOB & RMSE train & RMSE OOB \\
\hline
1 & 0 & 1 & 0 & 1 & 1 & 33.22 & 50.70 & 40.70 & 59.00 \\
0 & 0 & 1 & 0 & 1 & 1 & 32.93 & 48.63 & 43.14 & 59.22 \\
0 & 1 & 1 & 0 & 1 & 1 & 28.53 & 49.08 & 37.71 & 60.06 \\
1 & 1 & 1 & 0 & 1 & 1 & 29.56 & 52.38 & 36.88 & 61.11 \\
0 & 1 & 0 & 0 & 1 & 1 & 42.05 & 59.45 & 50.77 & 71.62 \\
1 & 0 & 0 & 0 & 1 & 1 & 46.58 & 64.07 & 55.77 & 72.66 \\
1 & 1 & 0 & 0 & 1 & 0 & 48.76 & 69.56 & 58.09 & 78.79 \\
1 & 1 & 0 & 0 & 1 & 1 & 40.73 & 67.49 & 49.49 & 79.86 \\
1 & 1 & 1 & 0 & 0 & 0 & 51.76 & 72.80 & 60.22 & 82.24 \\
1 & 1 & 0 & 0 & 0 & 1 & 52.97 & 74.30 & 59.95 & 82.75 \\
0 & 1 & 1 & 0 & 1 & 0 & 47.86 & 70.89 & 57.13 & 83.18 \\
1 & 0 & 1 & 0 & 0 & 1 & 51.42 & 74.36 & 59.34 & 83.19 \\
1 & 0 & 1 & 0 & 1 & 0 & 47.90 & 75.76 & 57.53 & 87.26 \\
1 & 1 & 1 & 0 & 0 & 1 & 51.41 & 79.30 & 58.70 & 87.61 \\
0 & 1 & 1 & 0 & 0 & 1 & 52.61 & 79.41 & 60.32 & 88.86 \\
1 & 1 & 1 & 0 & 1 & 0 & 44.69 & 75.01 & 53.82 & 89.12 \\
\hline
\end{tabular}
\end{table}

\begin{figure}[b]
    \includegraphics[width=\textwidth]{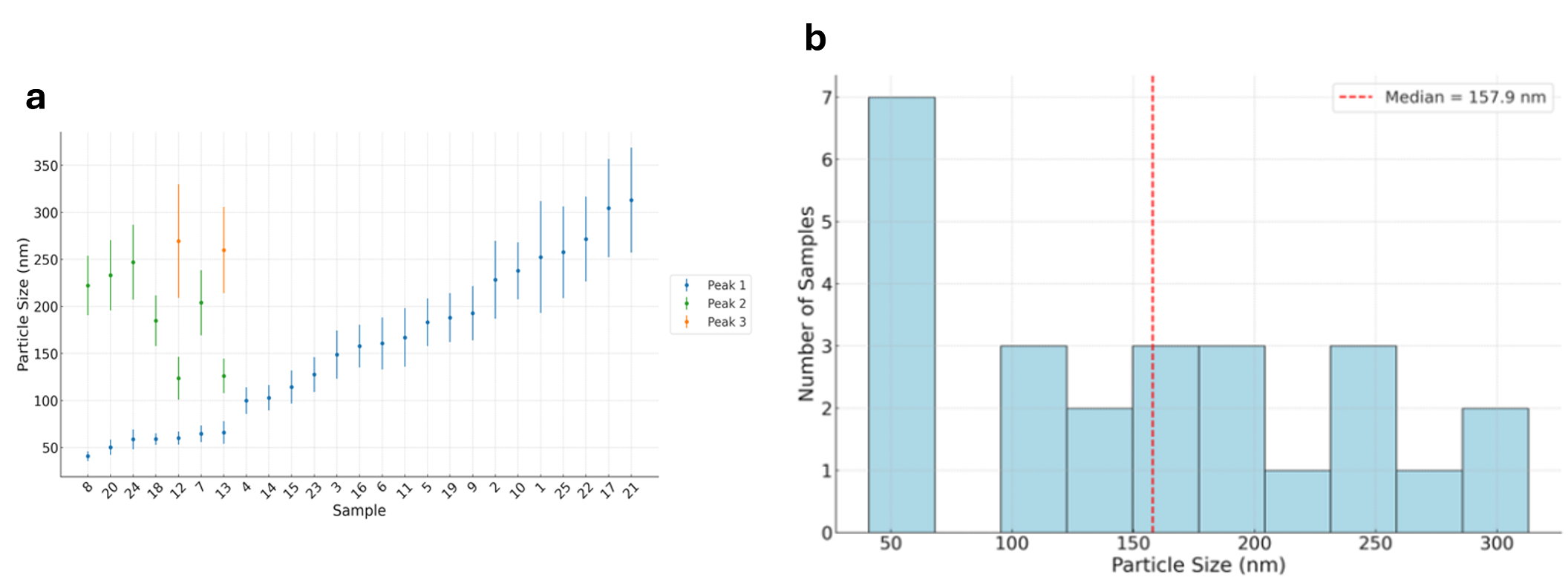}
    \caption{ a) DLS results showing particle hydrodynamic diameter and standard deviations for all detected peaks per sample. Each point represents a population of a distinct size (Peak 1 in blue, Peak 2 in green, and Peak 3 in orange), with error bars indicating the corresponding standard deviation. Samples are sorted in ascending order based on the size of Peak 1. Peaks above 5000 nm were excluded due to their negligible intensity ($<$1.5\%); b) Distribution of the smallest detected particle size per sample as measured by DLS. The red line indicates the median particle hydrodynamic diameter.\label{Fig:Particle sizes}}
\end{figure}

\begin{figure}[b]
    \includegraphics[width=0.5\textwidth]{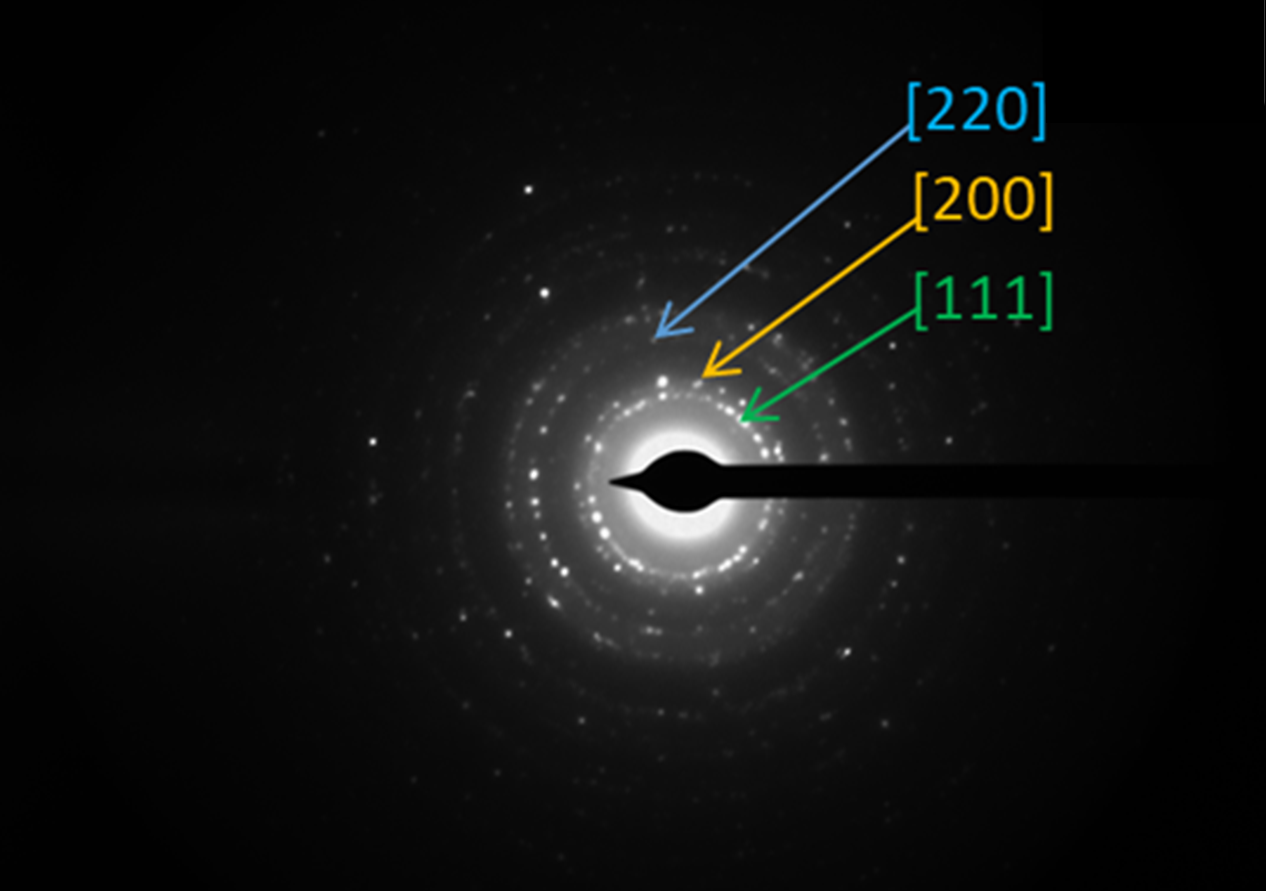}
    \caption{ Selected area electron diffraction pattern of $Cu^0$ nanoparticles from sample 22, recorded at an acceleration voltage of 120 kV. The presence of concentric diffraction rings indicates a polycrystalline material. The rings are indexed to the face-centred cubic (FCC) crystal lattice of metallic copper, with corresponding (111), (200) and (220) planes. The measured d-spacings for these planes are 2.08 \AA \space, 1.83 \AA \space, and 1.29 \AA \space, respectively, in good agreement with standard reference values.\label{Fig:SAED}}
\end{figure}

\begin{figure}[b]
    \includegraphics[width=1\textwidth]{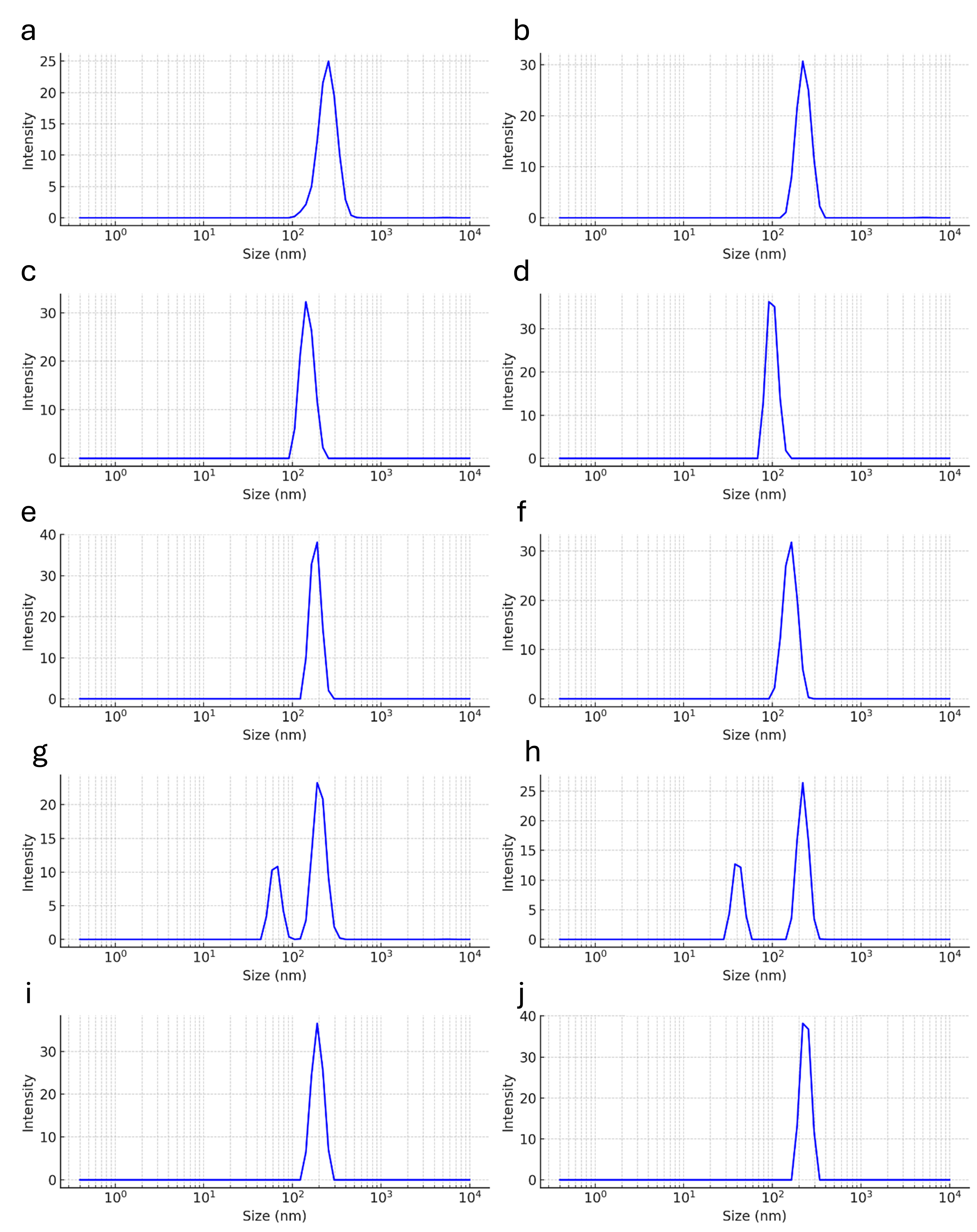}
    \caption{ Particle-hydrodynamic diameter distribution resulting from DLS of experimental synthesis a)Sample\_1; b)Sample\_2; c)Sample\_3; d)Sample\_4; e)Sample\_5; f)Sample\_6; g)Sample\_7; h)Sample\_8; i)Sample\_9; j)Sample\_10.\label{Fig:DLS reeks 1}}
\end{figure}

\begin{figure}[b]
    \includegraphics[width=1\textwidth]{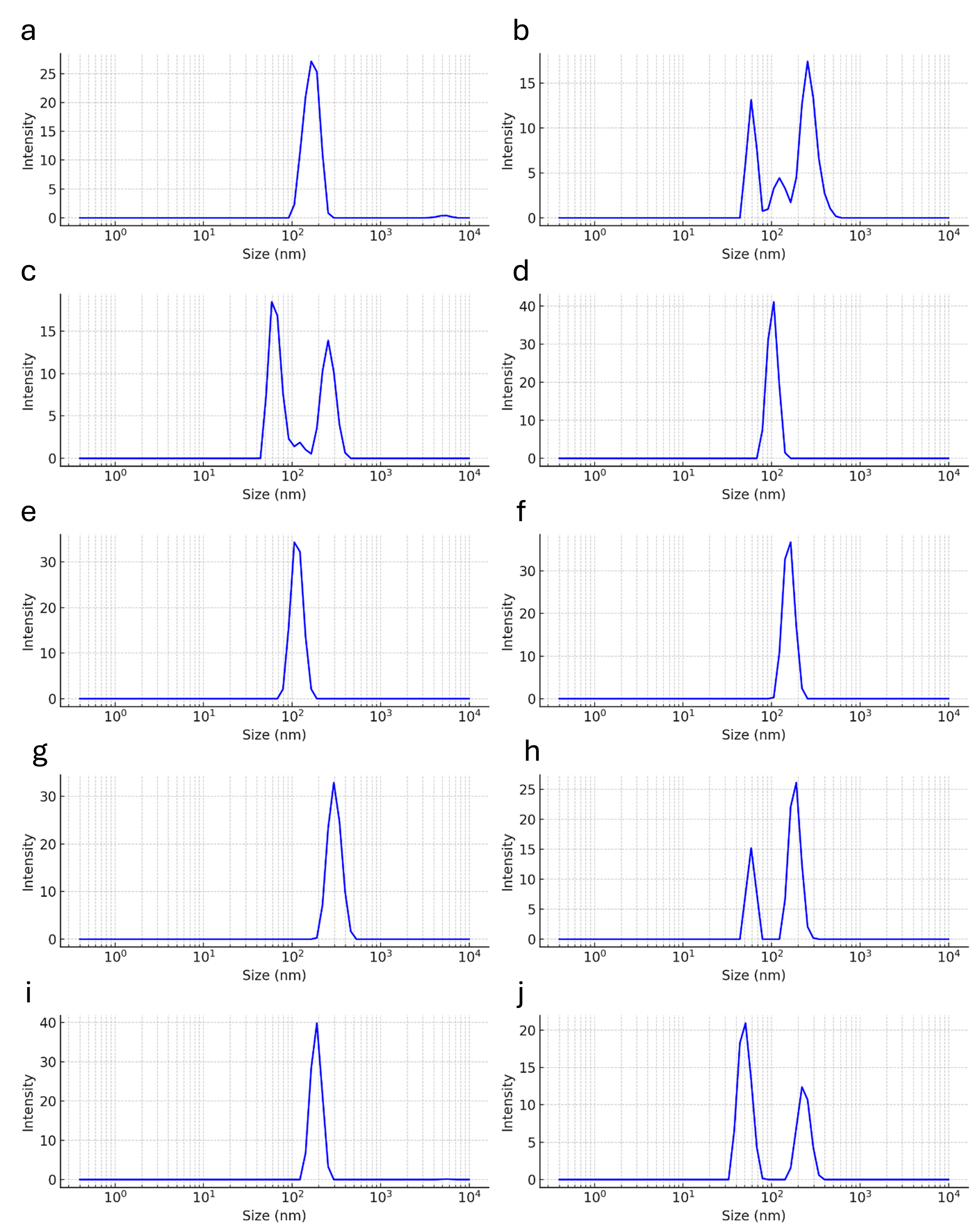}
    \caption{ Particle-hydrodynamic diameter distribution resulting from DLS of experimental synthesis a)Sample\_11; b)Sample\_12; c)Sample\_13; d)Sample\_14; e)Sample\_15; f)Sample\_16; g)Sample\_17; h)Sample\_18; i)Sample\_19; j)Sample\_20.\label{Fig:DLS reeks 2}}
\end{figure}

\begin{figure}[b]
    \includegraphics[width=0.97\textwidth]{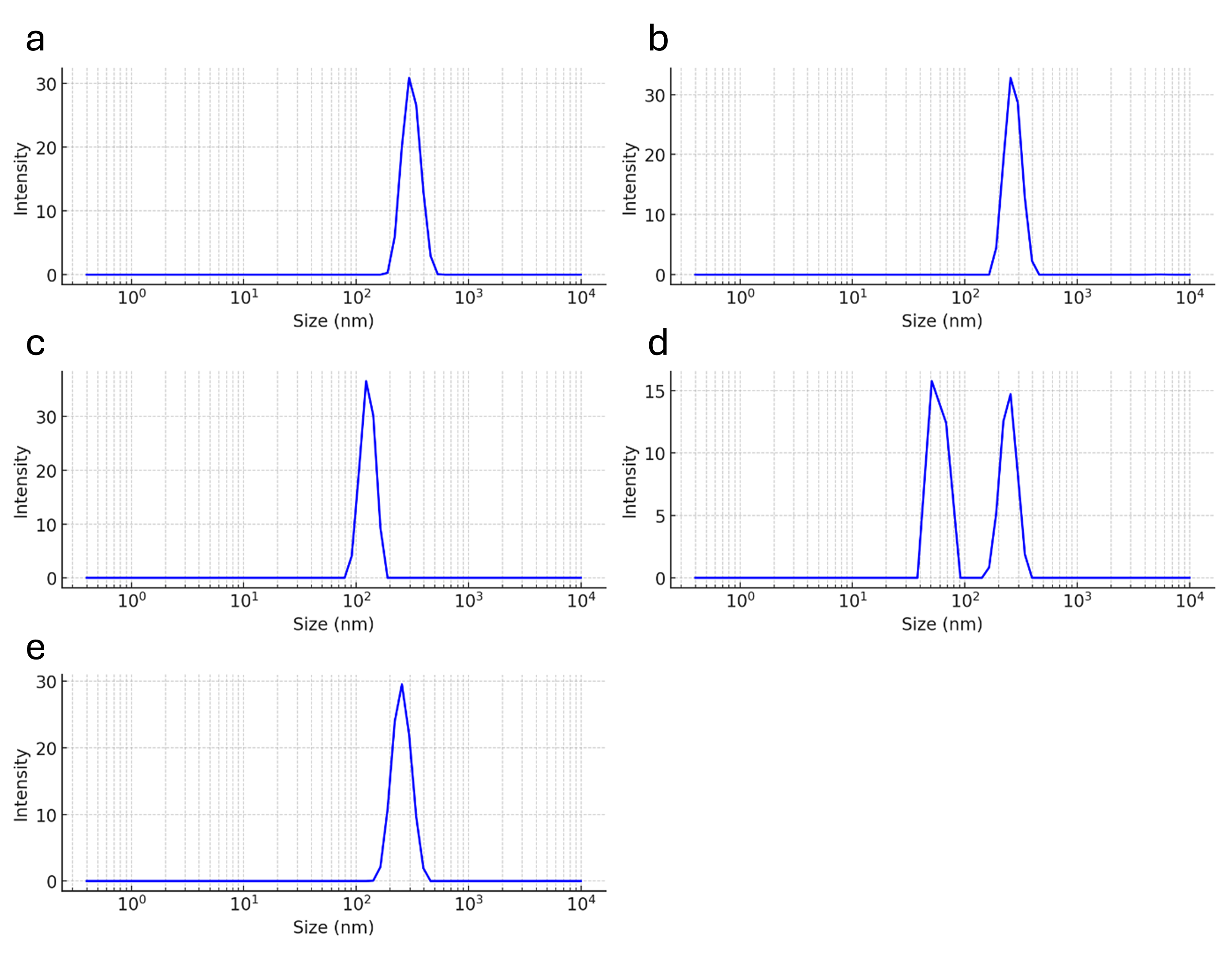}
    \caption{ Particle-hydrodynamic diameter distribution resulting from DLS of experimental synthesis a)Sample\_21; b)Sample\_22; c)Sample\_23; d)Sample\_24; e)Sample\_25.\label{Fig:DLS reeks 3}}
\end{figure}

\begin{figure}[b]
    \includegraphics[width=0.97\textwidth]{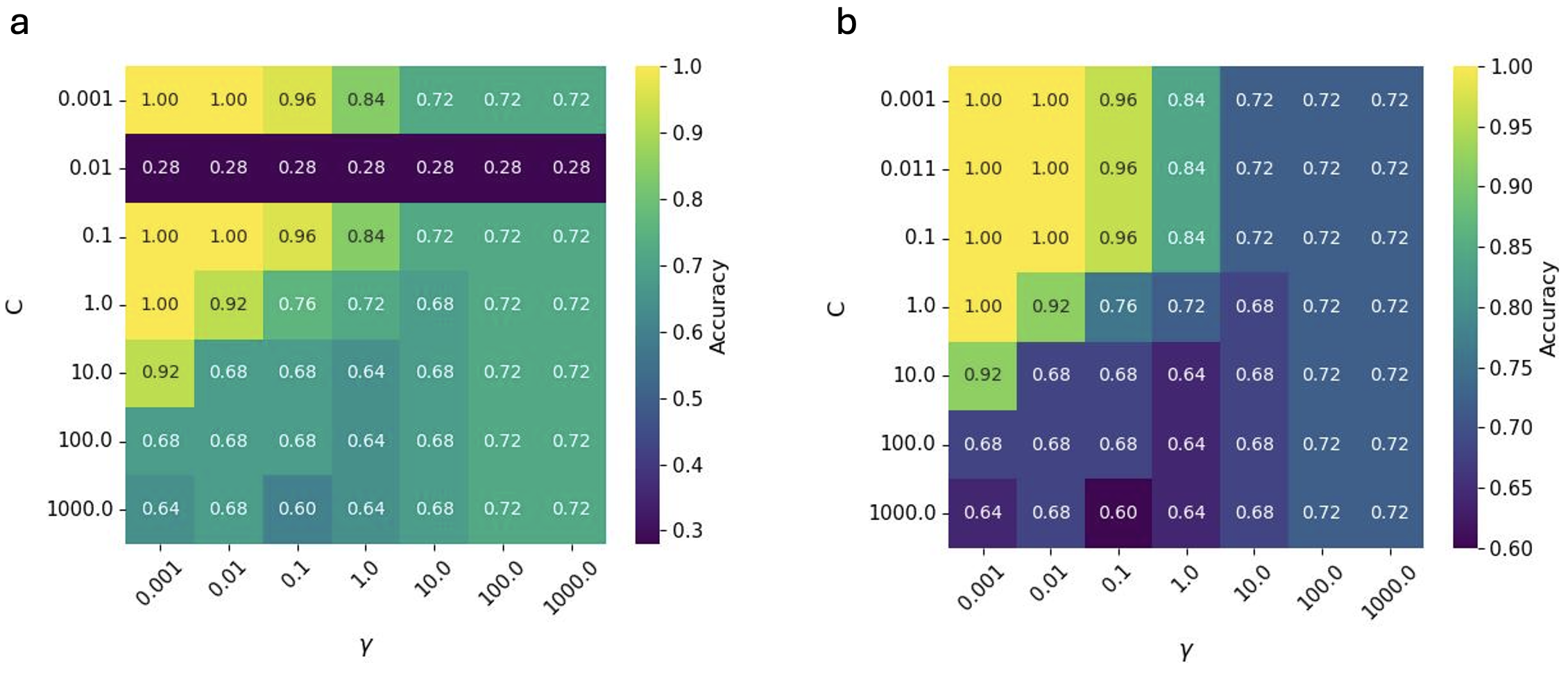}
    \caption{ Hyperparameter tuning of the SVM classifier. Heatmap showing the mean cross-validated accuracy as a function of the regularization parameter C and kernel parameter $\gamma$. a) Shows the full parameter grid, where the row corresponding to C $= 0.01$ yields nearly constant accuracy values across all $\gamma$, indicating degenerate model behavior. b) Shows a slightly adjusted parameter grid with C $= 0.011$, demonstrating that this effect is not intrinsic to the model but arises from algorithmic instability associated with the limited dataset size. This highlights the importance of careful interpretation of hyperparameter optimization results in small-data regimes. Notably, the selected optimal hyperparameters (C $= 1.0$, $\gamma = 1.0$) lie outside this regime, and therefore this behavior does not affect the final model selection.\label{Fig:SVM hyper tuning}}
\end{figure}

\begin{figure}[b]
    \includegraphics[width=0.5\textwidth]{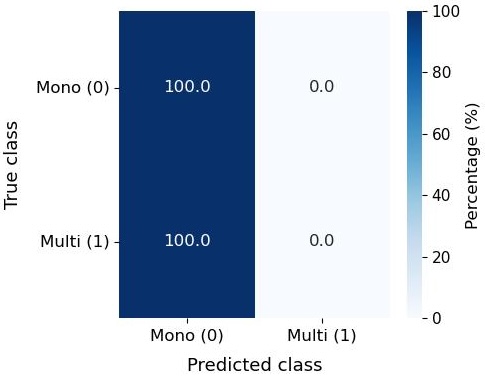}
    \caption{ Confusion matrix for an ensemble SVM model with hyperparameters C $= 0.1$ and $\gamma = 0.01$ yielding an apparent LOOCV accuracy of $1.0$. The ensemble model predicts only the mono-modal class, indicating a collapse to a single-class prediction despite the high LOOCV accuracy.\label{Fig:Confusion SVM Alternative Hyper}}
\end{figure}

\begin{figure}[b]
    \includegraphics[width=0.4\textwidth]{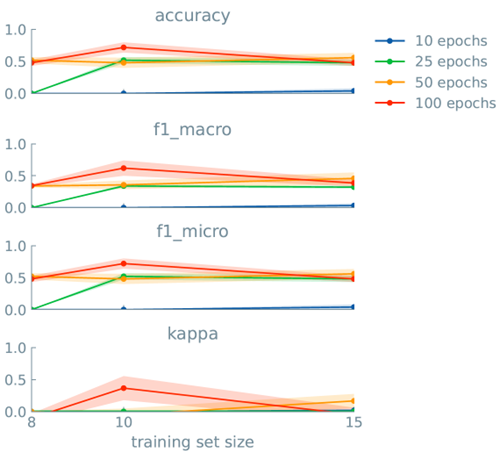}
    \caption{ Learning curve analysis of predictions for the binary class of the size of nanoparticles. Models fine-tuned with 10 (blue), 25 (green), 50 (orange) and 100 (red) epochs were validated. GPT-J was used as the base model.\label{Fig:Learning curve LLMs}}
\end{figure}

\begin{figure}[b]
    \includegraphics[width=0.9\textwidth]{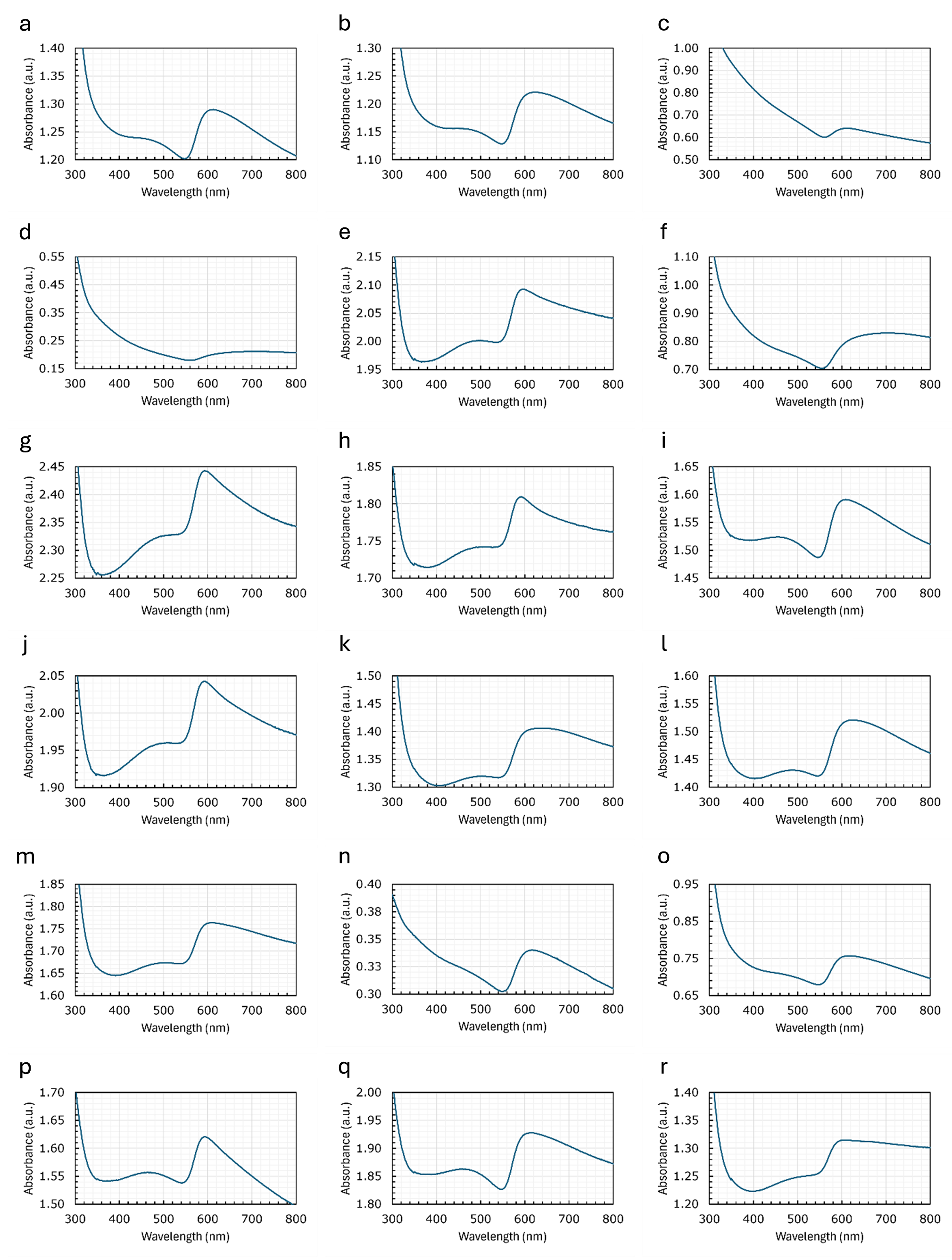}
    \caption{ UV-Vis spectra of a)Sample\_1; b)Sample\_2; c)Sample\_3; d)Sample\_4; e)Sample\_5; f)Sample\_6; g)Sample\_7; h)Sample\_8; i)Sample\_9; j)Sample\_10; k)Sample\_11; l)Sample\_12; m)Sample\_13; n)Sample\_14; o)Sample\_15;  p)Sample\_16; q)Sample\_17; r)Sample\_18.\label{Fig:UVvis_reeks1}}
\end{figure}

\begin{figure}[b]
    \includegraphics[width=0.9\textwidth]{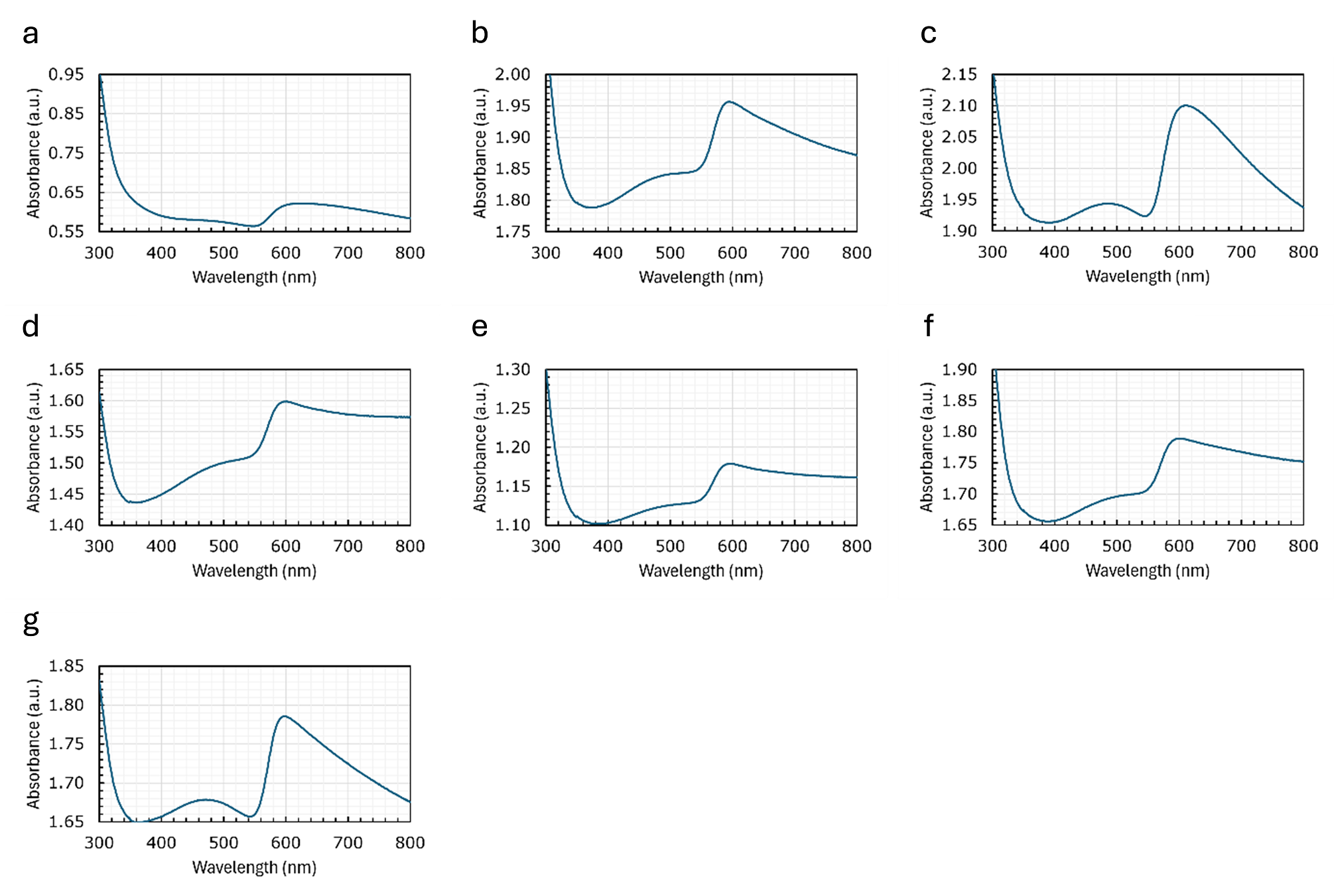}
    \caption{ UV-Vis spectra of a)Sample\_19; b)Sample\_20; c)Sample\_21; d)Sample\_22; e)Sample\_23; f)Sample\_24; g)Sample\_25.\label{Fig:UVvis_reeks2}}
\end{figure}

\begin{figure}[b]
    \includegraphics[width=1\textwidth]{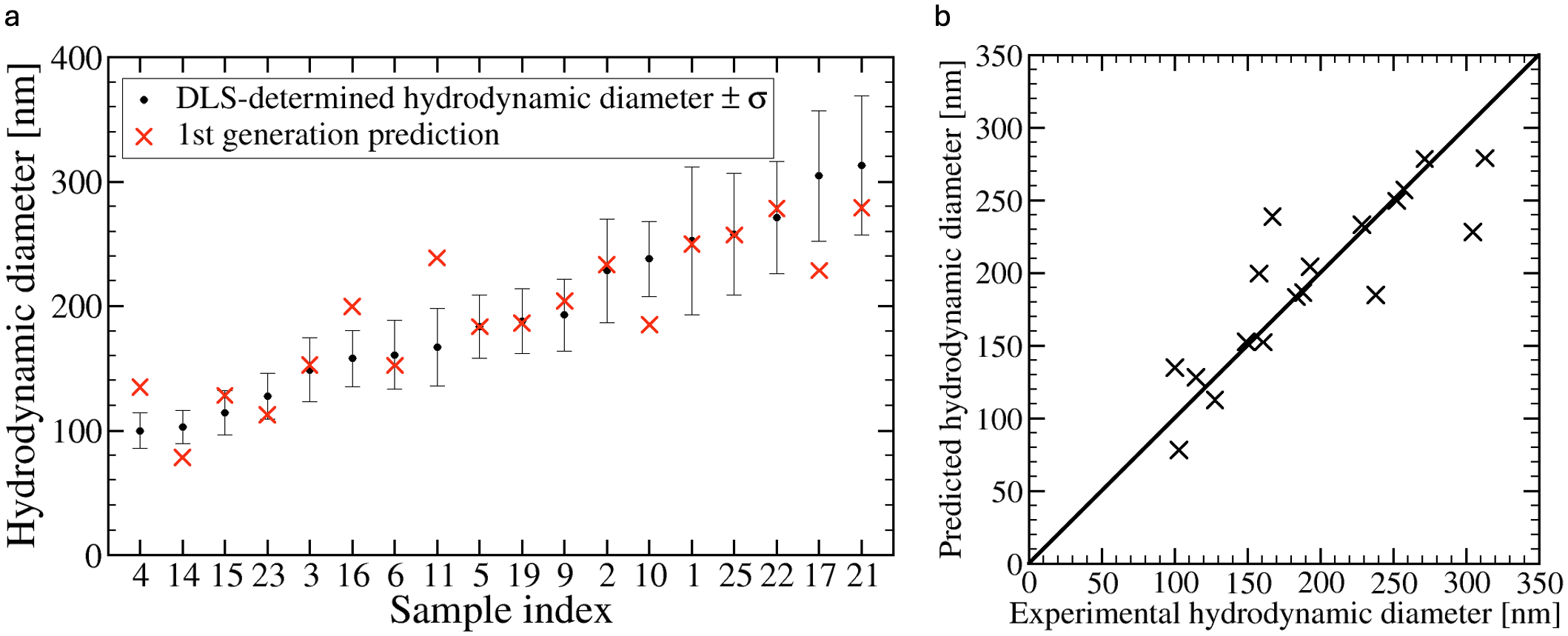}
    \caption{ a) Plot of the experimentally measured  hydrodynamic diameter and their standard deviation in ascending order (black) with the corresponding predicted particle size of the first generation ensemble model (red); b) Parity plot of the actual, experimental size against the corresponding predicted hydrodynamic diameter using the first generation ensemble model.\label{Fig:1st_generation}}
\end{figure}

\begin{figure}[b]
    \includegraphics[width=1\textwidth]{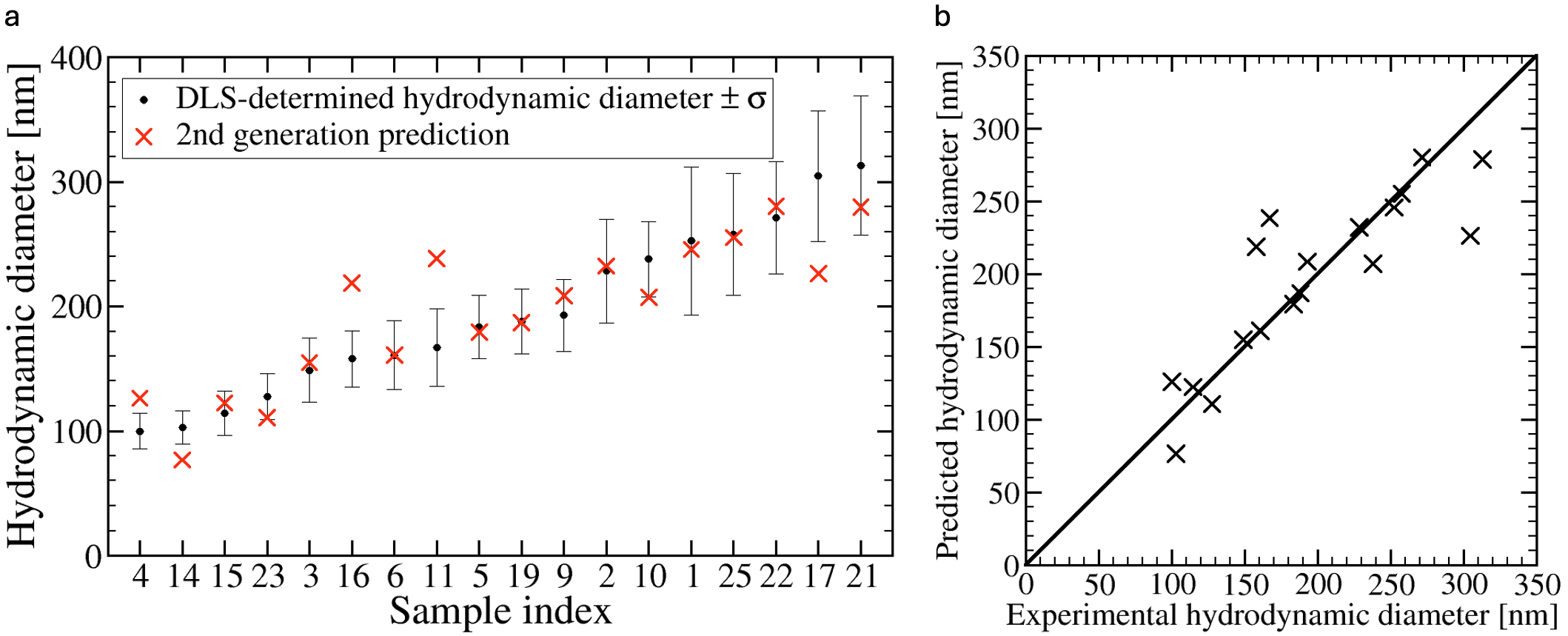}
    \caption{ a) Plot of the experimentally measured hydrodynamic diameter and their standard deviation in ascending order (black) with the corresponding predicted particle size of the second generation ensemble model (red); b) Parity plot of the actual, experimental hydrodynamic diameter against the corresponding predicted size using the second generation ensemble model.\label{Fig:2nd_generation}}
\end{figure}

\begin{figure}[b]
    \includegraphics[width=1\textwidth]{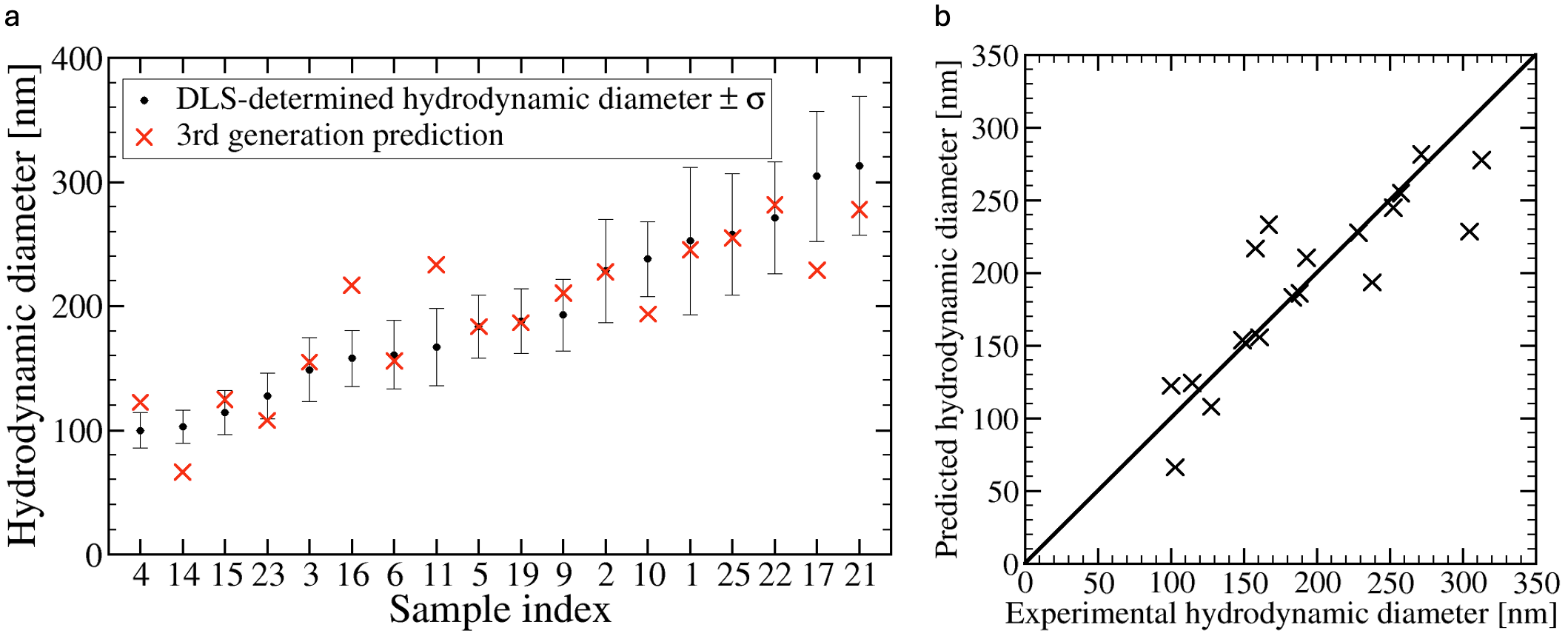}
    \caption{ a) Plot of the experimentally measured hydrodynamic diameter and their standard deviation in ascending order (black) with the corresponding predicted particle size of the third generation ensemble model (red); b) Parity plot of the actual, experimental hydrodynamic diameter against the corresponding predicted size using the third generation ensemble model.\label{Fig:3rd_generation}}
\end{figure}

\begin{figure}[b]
    \includegraphics[width=0.7\textwidth]{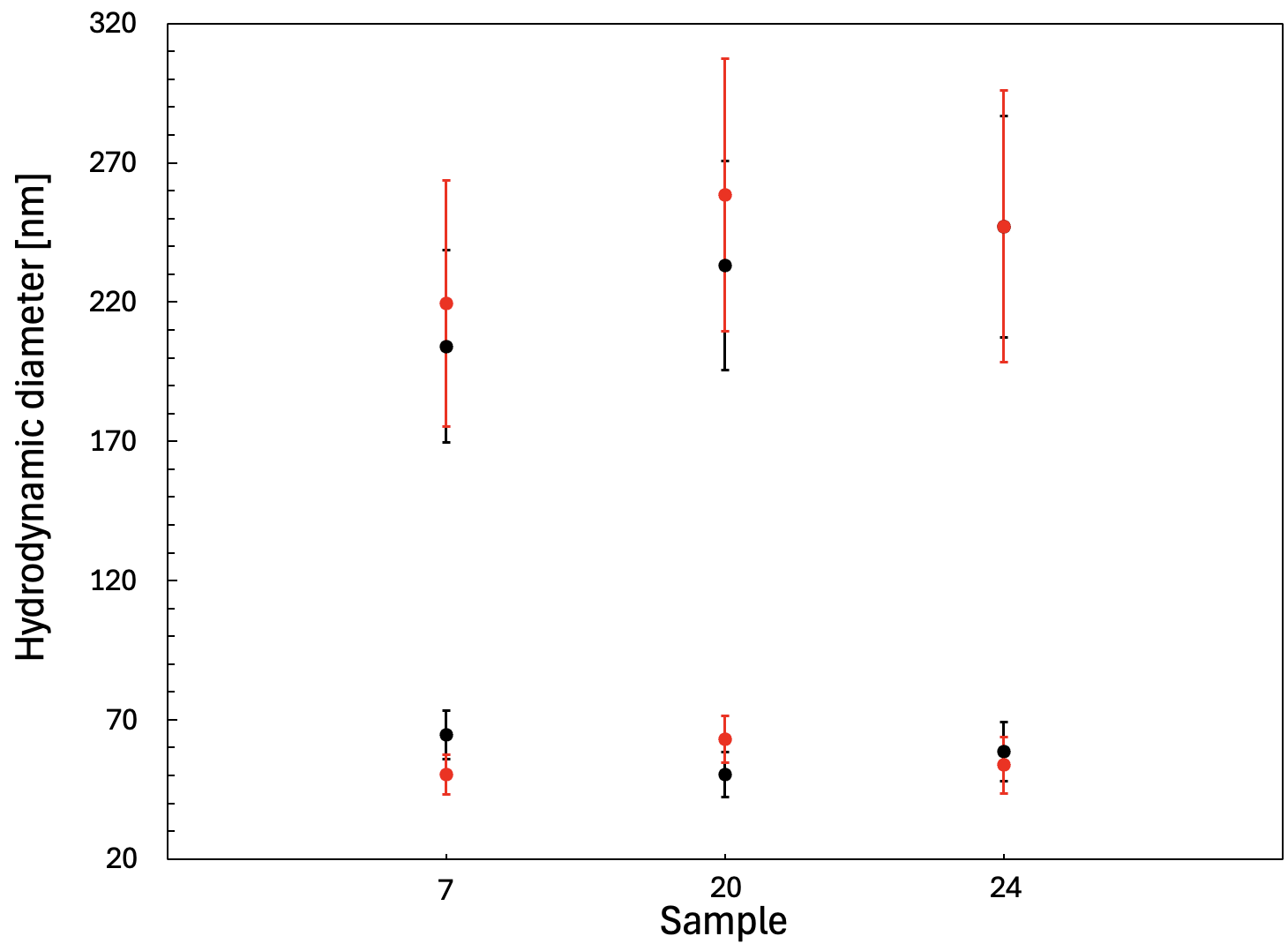}
    \caption{ Comparison of the DLS-derived hydrodynamic diameters obtained from the original syntheses (black) and independent repeat syntheses (red) for samples 7, 20, and 24. Error bars indicate the standard deviation of the DLS measurements.\label{Fig:repeatruns}}
\end{figure}

\begin{table}[b]
\centering
\caption{ Synthesis suggestions made by fourth generation model.
(a) Effective concentration of Cu(OAc)$_2$ precursor in 5 ml ethylene glycol in mM;
(b) Volume of 2.75 M TMAH in H$_2$O added to 5 ml ethylene glycol in $\mu$l;
(c) Time in minutes reaction mixture is kept on defined temperature;
(d) Temperature in $^\circ$C of reaction mixture for defined reaction time;
(e) Targeted hydrodynamic diameter in nm of the Cu NPs measured using DLS in ethanol;
(f) Obtained hydrodynamic diameter in nm of the Cu NPs and the $\sigma$ measured using DLS in ethanol.}
\label{tab:S1}
\begin{tabular}{ccccccc}
\hline
Sample & Cu conc.$^{a}$ [mM] & TMAH$^{b}$ [$\mu$l] & $Time^{c}$ [min] & Temp.$^{d}$ [$^\circ$C] & Targeted size$^{e}$ [nm] & Obtained size$^{f}$ [nm]\\
\hline
26 & 8.82 & 64.0 & 18 & 175 & 115 & 70.78 ($\pm$4.68) \\
27 & 2.30 & 16.8 & 4 & 195 & 115 & 117.90 ($\pm$32.06) \\
28 & 7.51 & 54.6 & 16 & 175 & 220 & 235.60 ($\pm$34.57) \\
29 & 3.91 & 27.8 & 7 & 189 & 220 & 128.60 ($\pm$49.45) \\
30 & 5.51 & 40.0 & 15 & 175 & 275 & 165.10 ($\pm$41.04) \\
31 & 6.01 & 43.6 & 4 & 179 & 275 & 189.60 ($\pm$59.50) \\
\hline
\end{tabular}
\end{table}

\begin{table}[tb]
\centering
\caption{ Synthesis suggestions made by DoE model.
(a) Targeted hydrodynamic diameter in nm of the Cu NPs measured using DLS in ethanol;
(b) Effective concentration of Cu(OAc)$_2$ precursor in 5 ml ethylene glycol in mM.}
\label{tab:S1}
\begin{tabular}{ccccccc}
\hline
Targeted size$^{a}$ [nm] & Cu conc.$^{b}$ [mM] \\
\hline
115 & 2.01 \\
115 & 10.10 \\
220 & 3.94 \\
220 & 8.16 \\
275 & No real solution \\
275 & No real solution \\
\hline
\end{tabular}
\end{table}

\end{document}